\documentclass[12pt]{article}
\usepackage{epsf}

\def\r#1{(\ref{#1})}
\def\Lms{\Lambda_{\overline{MS}}} 
\def\als{\alpha_s}
\newcommand{\nn}{\nonumber}
\newcommand{\mt}{m_{t}}
\newcommand{\mtb}{\overline{m}_{t}}

\newcommand{\mc}{m_{c}}
\newcommand{\ms}{m_{s}}
\newcommand{\md}{m_{d}}

\newcommand{\gev}{\, {\rm GeV}}

\newcommand{\bea}{\begin{eqnarray}}
\newcommand{\eea}{\end{eqnarray}}
\newcommand{\be}{\begin{equation}}
\newcommand{\ee}{\end{equation}}
\newcommand{\bi}{\begin{itemize}}
\newcommand{\ei}{\end{itemize}}

\newcommand{\epp}{\varepsilon^\prime/\varepsilon}
\newcommand{\ep}{\varepsilon_K}
\def\Im{\mathop{\mbox{Im}}}
\def\Re{\mathop{\mbox{Re}}}

\newcommand{\klpn}{K_L \to \pi^0 \nu \bar \nu}
\newcommand{\kppn}{K^+ \to \pi^+ \nu \bar \nu}
\newcommand{\kmm}{K_L \to \mu^+ \mu^-}
\newcommand{\kpe}{K_L \to \pi^0 e^+ e^-}
\newcommand{\sdz}{\bar s d Z}

\newcommand{\beq}{\begin{equation}}
\newcommand{\eeq}{\end{equation}}
\newcommand{\ba}{\begin{array}}
\newcommand{\ea}{\end{array}} 
\newcommand{\beqa}{\begin{eqnarray}}
\newcommand{\eeqa}{\end{eqnarray}}

\newcommand{\cH}{{\cal H}}

\newcommand{\cO}{{\cal O}}

\newcommand{\lsim}{\stackrel{<}{_\sim}}
\newcommand{\gsim}{\stackrel{>}{_\sim}}
\newcommand{\eps}{\varepsilon}
\newcommand{\epsp}{\varepsilon'}

\newcommand{\tq}{{\tilde q}}

\newcommand{\td}{{\tilde d}}
\newcommand{\tu}{{\tilde u}}
\newcommand{\tg}{{\tilde g}}
\newcommand{\ts}{{\tilde s}}
\newcommand{\ttop}{{\tilde t}}
\newcommand{\tlambda}{{\tilde \lambda}}
\newcommand{\deulr}{(\delta^{U}_{LR})}

\newcommand{\SM}{{\rm SM}}

\def\npb#1#2#3{    {\it Nucl. Phys. }{\bf B #1} (19#2) #3}
\def\plb#1#2#3{    {\it Phys. Lett. }{\bf B #1} (19#2) #3}
\def\prd#1#2#3{    {\it Phys. Rev. }{\bf D #1} (19#2) #3}

\def\prl#1#2#3{    {\it Phys. Rev. Lett. }{\bf #1} (19#2) #3}

\def\rmp#1#2#3{    {\it Rev. Mod. Phys. }{\bf #1} (19#2) #3}

\def\epjc#1#2#3{   {\it Eur. Phys. J. }{\bf C #1} (19#2) #3}

\def\jhep#1#2#3{   {\it JHEP  }{\bf #1} (19#2) #3}

\begin{document}
\setcounter{page}{0}
\thispagestyle{empty}
\begin{flushright}
{\small 
\baselineskip 0.5 truecm 
 TUM-HEP-353/99 \\
 ZU-TH 19/99    \\
 LNF-99/021(P)  \\
 OUTP--99--38P    \\  
 August  1999

}
\end{flushright}
\vskip 0.5 truecm 
 \centerline{\Large\bf Connections between $\epp$ and }
 \centerline{\Large\bf Rare Kaon Decays in Supersymmetry} 
\vskip1truecm
\centerline{
\bf A.J. Buras$^1$, G. Colangelo$^2$, G. Isidori$^3$,}
\centerline{
\bf A. Romanino$^4$ and L. Silvestrini$^1$ }
\bigskip

{\small 
\baselineskip 0.5 truecm 

\begin{center} 
{\sl 
${}^{1)}$ Technische Universit{\"a}t M{\"u}nchen, Physik Department \\
D-85747 Garching, Germany \\[3pt]
${}^{2)}$ Institut f{\"u}r Theoretische 
Physik der Universit{\"a}t Z{\"u}rich,\\
Winterthurerstr. 190, CH--8057 Z{\"u}rich--Irchel, Switzerland\\[3pt]
${}^{3)}$ INFN, Laboratori Nazionali di Frascati \\
P.O. Box 13, I--00044 Frascati, Italy\\[3pt]
${}^{4)}$ Department of Physics, Theoretical Physics, University of Oxford, \\ 
Oxford OX1\hspace{0.2em}3NP, UK\\[3pt]
} 
\end{center}
\vskip 0.8 truecm
\centerline{\bf Abstract}  
\vskip 0.3 truecm
We analyze the rare kaon decays $\klpn$, $\kppn$, $\kpe$ and $\kmm$ 
in conjunction with the CP violating ratio $\epp$ in a general class
of supersymmetric models in which $Z$- and magnetic-penguin
contributions can be substantially larger than in the Standard Model. 
We point out that radiative effects relate the double left-right mass
insertion to the single left-left one, and that the phenomenological
constraints on the latter reflect into a stringent bound on the 
supersymmetric contribution to the $Z$ penguin.
Using this bound, and those coming from recent data on $\epp$, 
we find
${\rm BR}(\klpn)\lsim 1.2\cdot 10^{-10}$, 
${\rm BR}(\kppn)\lsim 1.7\cdot 10^{-10}$, 
${\rm BR}(\kpe)_{\rm dir}\lsim 2.0\cdot 10^{-11}$, 
assuming the usual determination of the CKM parameters
and neglecting the possibility of  cancellations among 
different supersymmetric effects in $\epp$.
Larger values are possible, in principle, but rather unlikely. 
We stress the importance of a measurement of these three branching 
ratios, together with improved data and improved 
theory of $\epp$, in order to 
shed light on the realization of various 
supersymmetric scenarios.
We reemphasize that the most natural
enhancement of $\epp$, within supersymmetric models,
comes from chromomagnetic penguins and show that in this case sizable 
enhancements of $BR(\kpe)_{\rm dir}$ can also be expected.

}

\vfill 
\newpage
\section{Introduction}
\label{sec:intro}

Flavour-Changing Neutral Current (FCNC) processes provide a powerful
tool for testing the Standard Model and the physics beyond it. Of
particular interest are the rare kaon decays $\klpn$, $\kppn$ and $\kpe$
which are governed by $Z$-penguin diagrams. 
The latter diagrams play also a substantial role in the CP violating
ratio $\epp$. The most recent experimental results for this ratio, 
\begin{equation}\label{epeexp}
\Re(\epp) =\left\{ \begin{array}{ll}
 (28.0 \pm 4.1)\cdot 10^{-4} & {\rm (KTeV)\ \cite{KTEV}} \\
 (18.5 \pm 7.3)\cdot 10^{-4}  & {\rm (NA48)\ \cite{NA48}} 
 \end{array} \right.
\end{equation}
are in the ball park of the earlier result of the NA31 collaboration at
CERN, $(23.0 \pm 6.5)\cdot 10^{-4}$ \cite{barr:93}, and substantially
higher than the value of E731 at Fermilab, $(7.4 \pm 5.9)\cdot 10^{-4}$
\cite{gibbons:93}.
The grand average (according to the PDG recipe) including NA31, E731, KTeV
and NA48 results, reads 
\be
\Re(\epp) = (21.2\pm 4.6)\cdot 10^{-4}~,
\label{ga}
\ee
very close to the NA31 result but with a smaller error. The error should be
further reduced once complete data from both collaborations will be
analyzed. It is also of great interest to see what value for $\epp$ will be
measured by KLOE at Frascati, which uses a different experimental technique
than KTeV and NA48.

The estimates of $\epp$ within the Standard Model (SM) are generally below
the data but in view of large theoretical uncertainties stemming from
hadronic matrix elements one cannot firmly conclude that the data on $\epp$
imply new physics \cite{EP99,Dortmund,Rome,Trieste,BELKOV}. On the other
hand the apparent discrepancy between the SM estimates and the data invites
for speculations about non-standard contributions to $\epp$. Indeed the
KTeV result prompted several recent analyses of $\epp$ within various
extensions of the Standard Model (see e.g.~\cite{Sanda})
and particularly within supersymmetry \cite{Masiero,Babu}. 
Unfortunately these extensions have many parameters and if 
only $\epp$ is considered the analyses are not very conclusive.

The approach we want to pursue in the present paper is different:
we will adopt a model-independent point of view within a
generic supersymmetric extension of the Standard Model with minimal
particle content, and study what are the implications of a supersymmetric
$\epp$ for the rare decays. To do so we will use the mass-insertion
approximation \cite{HKR}.  Despite the presence of a large number of
parameters within this framework, only a few of them are allowed to
contribute substantially to $\epp$. Phenomenological constraints, coming
mainly from $\Delta S=2$ transitions \cite{GGMS}, make the contribution of
most of them to $\Delta S=1$ amplitudes
very small compared to the Standard Model one.
The only parameters
which survive are the left-right mass insertions
contributing to the Wilson coefficients of $Z$- and magnetic-penguin
operators. As we will discuss below, 
the reason for this simplification is a dimensional one: 
these are the only two classes of operators of dimension less 
than six contributing to $\epp$. 
Supposing that the enhancement of the Wilson
coefficients of either of these two (or both) type of operators 
is responsible for
the observed value of $\epp$, a corresponding effect in the rare decays
should be observed. In what follows we will analyze in detail the relations
between the size of the effect in $\epp$ and those in the rare decays.

The same kind of logic was already followed by two of us in
\cite{BS98}. There, this kind of analysis was carried through under the
assumption that the dominant effect in $\Delta S=1$ transitions was only an
enhanced $\sdz$ vertex. This analysis was motivated by an observation
of another two of us \cite{CI} that the branching ratios of 
rare kaon decays 
could be considerably enhanced, in a generic supersymmetric model, by large
contributions to the effective $\sdz$ vertex due to a double
left-right mass insertion. This double mass insertion had not been included
in earlier analyses of rare kaon decays in supersymmetry
\cite{NIRWO,BRS}. In the latter papers only single mass insertions were
taken into account, leading to modest enhancements of rare-decay branching
ratios, up to factors 2-3 at most, as opposed to the possible enhancement
of more than one order of magnitude allowed by the double mass insertion
\cite{CI}. The conclusion of the analysis in \cite{BS98} was that the data
on $\epp$ may constrain considerably the double left-right mass insertion
and the corresponding enhancement of the rare-decays branching ratios.

In the present paper we will improve the analysis in \cite{BS98}
with the aim to answer the following questions:
\begin{itemize}
\item 
Can the large double mass insertions suggested in \cite{CI} be
further constrained? As we will see this is indeed the case.
\item 
What is the impact of these new constraints on the analysis in
\cite{BS98}?
\item 
What is the impact on this analysis of contributions from
chromomagnetic and $\gamma$-magnetic 
penguins to $\epp$ and $\kpe$ respectively?
\end{itemize}  
As we mentioned above, in generic supersymmetric theories a sizable
contribution to $\epp$ could also be generated by the chromomagnetic-dipole
operator. Actually, within supersymmetric models with approximate flavor
symmetries, the latter mechanism seems to be more natural than a strong
enhancement of the $\sdz$ vertex \cite{Masiero}. Interestingly, if
the Wilson coefficient of the chromomagnetic-dipole operator gets enhanced,
one should also expect a sizable effect in the branching ratio of
$K_L\to\pi^0e^+e^-$, due to the $\gamma$-magnetic penguin. In fact their
Wilson coefficients receive contributions from the same type of 
mass insertion.  

The paper is organized as follows: In Section 2 we identify the dominant
SUSY contributions to $|\Delta S|=1$ amplitudes as those of dimension less
than six. In Section 3 we summarize the effective Hamiltonian for $|\Delta
S|=1$ transitions concentrating on the operators of dimension four
(effective $\sdz$ vertex) and five (magnetic penguins) and their
corresponding Wilson coefficients. Here we introduce three effective
couplings which characterize the supersymmetric contributions to the Wilson
coefficients of these operators: $\Lambda_t$ for the $Z$ penguin and
$\Lambda_g^\pm$ for the magnetic ones. In Section 4 we collect the basic
formulae for $\epp$ and rare kaon decays in terms of these effective
couplings. In particular we calculate the magnetic contributions to $\epp$
and $\kpe$. In Section 5 we analyze indirect bounds on the effective
couplings. The main result of this section is an improved upper bound on
$|\Lambda_t|$ coming from renormalization group considerations. In Section
6 we present a detailed numerical analysis of rare kaon decays taking into
account the recent data on $\epp$, the present information on the short
distance contribution to ${\rm BR}(K_L\to\mu^+\mu^-)$ and the bounds on
effective couplings derived in Section 5. Analyzing various scenarios we
calculate upper limits on ${\rm BR}(\kpe)_{\rm dir}$, ${\rm BR}(\klpn)$ and
${\rm BR}(\kppn)$. We present a summary and our conclusions in Section 7.

\section{SUSY contributions to $|\Delta S|=1$ amplitudes}
In the Standard Model FCNC
amplitudes are generated only at the quantum level.
The same remains true also in low-energy supersymmetric models
with unbroken $R$ parity, minimal particle content and generic flavour
couplings.  
The flavour structure of a generic SUSY model is quite complicated
and a convenient way to parametrize the various flavour-mixing
terms is provided by the so-called mass-insertion approximation
\cite{HKR}. This consists in choosing a simple basis for the gauge 
interactions and, in that basis, to perform a 
perturbative expansion of the squark mass matrices
around their diagonal. In the following we will employ a 
squark basis where all quark-squark-gaugino vertices involving 
down-type  quarks are flavor diagonal.

In the case of $|\Delta S|=1$ transitions we can distinguish between two
large classes of one-loop diagrams: 
\begin{itemize}
\item{} 
{\it Box diagrams}. These are present both in $|\Delta S|=1$ and 
$|\Delta S|=2$ amplitudes. In both cases the 
integration of the heavy degrees of freedom, associated with the 
superpartners, necessarily leads to effective four-quark operators 
of dimension six. The Wilson coefficients of these operators 
are therefore suppressed by two powers of a supersymmetry-breaking scale,
that we generically denote by $M_S$. 
Here $1/M_S^2$ plays a role similar to $1/M^2_W$ in the SM case.

Since any mass-insertion carries at most $|\Delta S|=1$, 
the leading contribution to $|\Delta S|=2$ transitions 
starts at second order in this expansion. 
Denoting by $\delta$ the generic ratio of off-diagonal terms over 
diagonal ones in the squark mass matrices, 
the coupling of $|\Delta S|=2$ effective operators 
turns out to be of $\cO(\delta^2/M_S^2)$.
This has to be compared with the dominant SM coupling
that is of $\cO(\lambda_t^2/M_W^2)$, where 
$\lambda_t=V^*_{ts} V_{td}$. If we then
impose that the supersymmetric contribution 
to $|\Delta S|=2$ amplitudes is at most of the order of 
the SM one, we find 
\beq
\delta/M_S \lsim  \lambda_t /M_W~.
\label{DF2bound}
\eeq
In the case of $|\Delta S|=1$ amplitudes, the leading 
supersymmetric contribution starts already at first 
order in $\delta$, similarly to the SM one that is 
linear in $\lambda_t$.
However, the dimensional suppression factor is always $1/M_S^2$ in the 
SUSY case and $1/M_W^2$ in the SM one. 
Therefore, if $M_S \gg M_W$,
the constraint (\ref{DF2bound}) implies that the supersymmetric
contribution to $|\Delta S|=1$ box diagrams is suppressed  
with respect to the SM one.
This naive argument is confirmed by the 
detailed analysis of \cite{GGMS}, where it has 
been shown that $|\Delta S|=2$ constraints  
always dominate over $|\Delta S|=1$ ones,
as long as we consider only dimension-six operators
generated by box diagrams with gluino exchange.
\item{} 
{\it Penguin diagrams}. At the one-loop level
this kind of diagrams is present only in $|\Delta S|=1$
amplitudes. Effective operators with lowest dimension generated by 
photon and gluon penguins are the so-called ``magnetic'' 
operators of dimension five. The coupling of these operators is of 
$\cO(\delta/M_S)$ and therefore potentially competing 
with the SM contributions even if we impose the bound
(\ref{DF2bound}).
This naive conclusion is again confirmed by detailed analyses of
gluino mediated amplitudes \cite{GGMS}. In this 
context it is found that only the chromomagnetic 
operator, induced by $\td_{L(R)}-\ts_{R(L)}$
mixing, could lead to sizable $(\gsim 10^{-3})$
contributions to $\epsp/\eps$ without violating any
constraints from $\eps$.

A different situation occurs in the case of $Z$-penguin diagrams, where the
breaking of $SU(2)_L$ allows to build an effective dimension-four operator
of the type $s_L\gamma^\mu d_L Z_\mu$. Denoting by $C_Z$ the dimensionless
coupling of this operator, the integration of the heavy $Z$ field leads to
an effective four-fermion operator proportional to $C_Z/M_Z^2$ without any
explicit $1/M_S$ suppression. This potential enhancement is partially
compensated by the fact that the leading contribution to $C_Z$ arises only
at second order in the mass-insertion \cite{CI}. However, the absence of
any $1/M_S$ suppression makes this term particularly interesting both for
rare decays \cite{CI} and $\epsp/\eps$ \cite{BS98}.
\end{itemize}

\noindent
Given the above considerations, in the following we will restrict our
attention only to the dominant SUSY effects in $|\Delta S|=1$ amplitudes:
those generated by the ``magnetic'' dimension-five operators, induced by
gluino exchange, and those generated by the $\sdz$ vertex mediated by
chargino exchange.  Interestingly, under this assumption only the
off-diagonal left-right entries of squark mass matrices are involved, in
particular the $\td_{L(R)}-\ts_{R(L)}$ mixing for the magnetic operators
and the $\tu^{(s,d)}_{L}-\ttop_{R}$ one for the $\sdz$ vertex.

What we will not consider are the gluino 
and the chargino contributions  
to irreducible dimension-six operators. 
The former have been explicitly calculated in 
\cite{GGMS} and found to be negligible, the latter
are suppressed by $\cO(M^2_W/M^2_S)$ with respect 
to the corresponding contributions mediated by  
the $\sdz$ vertex. To control the accuracy of our 
approximation,  we have explicitly checked that the impact of 
these terms is below $10\%$, with respect to the 
dominant ones, for squark/gaugino masses above $\sim 300$~GeV.
Finally, we will completely ignore the 
neutralino contributions which are known to be 
negligible due to the smallness of both 
electroweak and  down-type Yukawa couplings \cite{BRS}.

Since a large $\sdz$ vertex is already present in
the SM, the corresponding SUSY corrections can be easily 
incorporated without modifying 
the structure of the SM $|\Delta S|=1$ effective Hamiltonian.
On the other hand, the dimension-five operators, 
neglected within the SM, require an adequate treatment and
will be discussed in detail below.

\section{Effective Hamiltonian}
\label{sect:Heff}
\subsection{Operators and Wilson Coefficients}
On the basis of the discussion in the previous section, we introduce here
the effective Hamiltonian containing all the relevant operators of
dimension smaller than six.  The only dimension-four operator of interest
is the one given by the $\sdz$ vertex:
\begin{equation}
  \label{eq:Wds}
  {\cal H}^{d=4}_{\rm eff} = -\frac{G_F}{\sqrt{2}} \frac{e}{ \pi^2} M_Z^2
  \frac{\cos \Theta_W}{\sin \Theta_W} Z_{ds} \bar s_L \gamma_\mu
   Z^\mu d_L \,+\, {\rm h.c.}~,
\end{equation}
where 
\begin{equation}
  \label{eq:Wsm}
  Z_{ds} = \lambda_t C_0(x_t)+\tilde\lambda_t H_0(x_{q \chi})~.
\end{equation}
Here the first term on the r.h.s is the Standard Model contribution
(evaluated in the 't Hooft-Feynman gauge)
and the second one represents the dominant supersymmetric effect.
The couplings $\lambda_t$ and $\tilde\lambda_t$ are defined by
\be\label{ll1}
\lambda_t = V_{ts}^* V_{td}~, \qquad 
\tlambda_t = \deulr_{23} \deulr_{13}^*~,
\ee
where $V_{ij}$ are the elements in the CKM matrix and,
denoting by $M^2_{[U,D]}$ the squark mass matrices,
\be\label{deltas}
\left(\delta^{[U,D]}_{AB}\right)_{ij}=\left(M^2_{[U,D]}\right)_{i_A j_B}
\left/ \langle M^2_{[U,D]} \rangle \right.~.
\ee
Explicit expressions for the functions $C_0$ and $H_0$ will be given below. 

The magnetic operators of dimension five appear in the effective 
Hamiltonian in the following way:
\begin{equation}\label{Heff5}
{\cal H}_{\rm eff}^{d=5} = (C^+_\gamma Q^+_\gamma + C^-_\gamma Q^-_\gamma 
+ C^+_g Q^+_g + C^-_g Q^-_g) + {\rm h.c.}~,
\end{equation}
where we have chosen the following operator basis:
\begin{eqnarray}
Q^\pm_\gamma&=&\frac{Q_d e}{16 \pi^2}
        \left( {\bar s}_L \sigma^{\mu \nu} F_{\mu\nu} d_R \pm 
               {\bar s}_R \sigma^{\mu \nu} F_{\mu\nu} d_L \right)~, \\
Q^\pm_g&=&\frac{g}{16 \pi^2}
        \left( {\bar s}_L \sigma^{\mu \nu} t^a G^a_{\mu\nu} d_R \pm
               {\bar s}_R \sigma^{\mu \nu} t^a G^a_{\mu\nu} d_L \right)~, 
\end{eqnarray}
Full expressions for the Wilson coefficients generated by gluino exchange 
at the SUSY scale can be found in
\cite{GGMS}. We are interested here only in the contributions
proportional to $1/m_{\tilde g}$, which are given by 
\begin{eqnarray}
\label{eq:cgamma}
C^\pm_\gamma(m_{\tilde g})&=& \frac{\pi \alpha_s(m_{\tilde
    g})}{m_{\tilde g} } \left[
  \left(\delta^{D}_{LR}\right)_{21} \pm 
  \left(\delta^{D}_{LR}\right)^*_{12}\right]  
  F_0(x_{g q}) \; \; , \\
C^\pm_g(m_{\tilde g})&=& \frac{\pi \alpha_s(m_{\tilde g})}{m_{\tilde g}}
 \left[ \left(\delta^{D}_{LR}\right)_{21} \pm
   \left(\delta^{D}_{LR}\right)^*_{12} \right] G_0(x_{g q})
 \; \; ,
\end{eqnarray}
where the $\delta_{ij}$ are defined in (\ref{deltas}) and
the functions $F_0$ and $G_0$ are given in (\ref{F0}) and (\ref{G0}).

In the $(Q^\pm_g, Q^\pm_\gamma)$ basis, 
the leading order anomalous dimension matrix reads
\begin{equation}
\gamma = \left(
\begin{array}{cc}
8/3 & 0 \\
& \\
32/3 & 4/3
\end{array}
\right)~.
\end{equation} 
Therefore, integrating out SUSY particles at the scale
$m_{\tilde g} > m_t$,  one has
\begin{eqnarray}
C^\pm_\gamma(m_c) &=& \eta^2 \left[ C^\pm_\gamma(m_{\tilde g}) + 8\,
  (1-\eta^{-1})\, C^\pm_g(m_{\tilde g}) \right], \\ 
C^\pm_g(m_c) &=& \eta\, C^\pm_g(m_{\tilde g}),
\end{eqnarray}
where
 \begin{equation}\label{eta}
\eta=\left(\frac{\alpha_s(m_{\tilde g})}{\alpha_s(m_t)}\right)^\frac{2}{21}
        \left(\frac{\alpha_s(m_t)}{\alpha_s(m_b)}\right)^\frac{2}{23}
        \left(\frac{\alpha_s(m_b)}{\alpha_s(m_c)}\right)^\frac{2}{25}~.
\end{equation}
The dimension-five operators in (\ref{Heff5}) in principle 
mix also with $Q_2$, the leading dimension-six operator of the 
SM $|\Delta S|=1$ effective Hamiltonian (see e.g. \cite{BBL}). 
However, the effect of this mixing can be neglected as long as we 
are interested in large enhancements of the Wilson coefficients
of the dimension-five operators with respect to the SM case (more 
than one order of magnitude in the imaginary parts, as suggested 
in \cite{Masiero}). Therefore, as first approximation, 
in the following we will neglect the mixing 
of $Q_{g(\gamma)}^\pm$ with $Q_2$.

\subsection{Basic Functions}
The basic functions relevant for our analysis are
\begin{equation}\label{B0}
B_0(x)={1\over
4}\left[{x\over{1-x}}+{{x\ln (x)}\over{(x-1)^2}}\right]~,
\end{equation}
\begin{equation}\label{C0}
C_0(x)={x\over 8}\left[{{x-6}\over{x-1}}+{{3x+2}
\over{(x-1)^2}}\;\ln (x) \right]~,
\end{equation}
\be\label{H0}
H_0(x) = - \frac{x(x^3-6x^2+3x+2+6 x \ln(x))}{48(1-x)^4}~,
\ee
\begin{eqnarray}
F_0(x) &=&
 {{4x(1 + 4\,x - 5\,{x^2} + 4\,x\,\ln (x) + 2\,{x^2}\,\ln (x))}\over 
    {3\,{{\left( 1 - x \right) }^4}}}~, \label{F0}\\
G_0(x) &=&
 \frac{x(22-20x-2x^2+16x\ln(x) -x^2\ln(x)+9\ln(x))}{3(1-x)^4}~,
\label{G0}
\end{eqnarray}
with the corresponding mass ratios
\be\label{xx}
 x_t = m_t^2/m_W^2 ~,\qquad
 x_{q\chi} = m_{\tq}^2/m_{\tilde \chi}^2~,  \qquad
  x_{gq} = m_{\tg}^2/m_\tq^2~.
 \ee
$B_0(x_t)$ and $C_0(x_t)$ are the box and $Z^0$ penguin diagram functions
in the Standard Model respectively.
The function $H_0(x_{q\chi})$ appears in the SUSY contribution to the 
$\sdz$ vertex \cite{CI}. The
functions $F_0(x_{gq})$ and $G_0(x_{gq})$ enter the contributions of
$\gamma$-magnetic and chromomagnetic penguin operators
respectively \cite{GGMS}. 

\subsection{Effective couplings}

\begin{figure}[t]
\begin{center}
\leavevmode
\epsfxsize=8cm\epsfbox{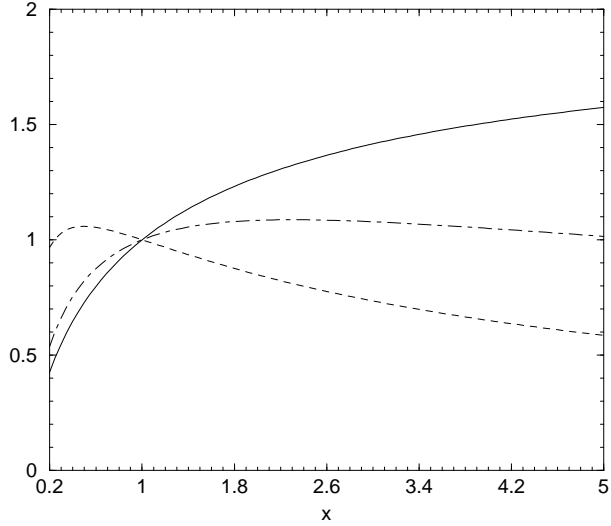}
\end{center}
\caption{\label{fig:hgxdep} Dependence on $x$ of the 
functions $H(x)/H(1)$ (solid),  $G(x)/G(1)$ (dashed),
$F(x)/F(1)$ (dot-dashed).}
\end{figure}

The SUSY Wilson coefficients which we have given above depend explicitly on 
the sparticle masses via the functions $H_0$, $F_0$ and $G_0$. The
dependence is not very strong, as can be seen from Fig.~\ref{fig:hgxdep},
where we plot the three functions normalized to their values at $x=1$
$(H_0(1)=-1/96,\ F_0(1)=2/9,\ G_0(1)=-5/18)$.
On the other hand the relations between $\epp$ and the rare decays which we
want to investigate here, are almost independent from the spectra of the SUSY
particles. In fact these relations are most conveniently described in terms
of three effective couplings defined as follows:
\bea
\label{LLdef}
\Lambda_t &=& \left[ \deulr_{23} \deulr_{13}^* \right] H_0(x_{q
  \chi}) \; \; , \nonumber \\ 
\Lambda^\pm_g &=& \left[ \left(\delta^{D}_{LR}\right)_{21} \pm
   \left(\delta^{D}_{LR}\right)^*_{12} \right] G_0(x_{g q})
 \; .
\eea
It is worthwhile to point out that most of the results 
presented in Section~\ref{sect:analysis}
are valid also if these couplings 
are defined in a more general way, starting from the Wilson 
coefficients of $Z$-penguin and chromomagnetic operators.
This way one could efficiently include also 
subleading contributions in the mass-insertion 
approximation. This is however beyond the scope 
of the present analysis.

\section{Basic Formulae for $\epp$ and Rare Decays}
\label{sec:BF}
In this section we collect the formulae for $\epp$ and rare K decays
which we have used in our analysis. These formulae can be considered as
the generalization of the corresponding expressions in \cite{BS98}
to include contributions of the chromomagnetic and $\gamma$-magnetic
operators to $\epp$ and $K_L\to\pi^0 e^+e^-$ respectively. However,
we stress that here we will treat the effective $\sdz$ vertex 
differently than in \cite{BS98}, separating explicitly SM and 
supersymmetric contributions as shown in (\ref{eq:Wds}).
The latter will be described in terms of the effective 
coupling $\Lambda_t$ defined in (\ref{LLdef}).

\subsection{Magnetic contributions to $\epp$ and $K_L\to\pi^0 e^+ e^-$}

The matrix elements of the magnetic operators $Q_{g,\gamma}^\pm$ between a
$K^0$ and an $n$-pion state are difficult to calculate. In the following we 
will normalize them by using the value obtained in model calculations, and 
introduce the corresponding $B$ factors which we will then vary inside our 
estimates of the uncertainties. We will use:
\bea
\label{eq:BG}
\langle (\pi\pi)_{I=0} | Q_g^- | K^0 \rangle &=& \sqrt{3 \over 2}
\frac{ 11}{16 \pi^2} \frac{\langle \bar q q \rangle}{F_\pi^3} m_\pi^2 \; B_G 
\; \; , \\
\label{eq:BT}
\langle \pi^0 | Q_\gamma^+ | K^0 \rangle &=& {Q_d e \over 16 \pi^2}
{i\sqrt{2} \over m_K} p_\pi^\mu p_K^\nu F_{\mu \nu} \; B_T 
\; \; , \\
\langle (\pi\pi)_{I=0} | Q_g^+ | K^0 \rangle &=&\langle \pi^0 | Q_\gamma^-
| K^0 \rangle \; = \; 0 \; \; .
\eea
For $B_G=1$ Eq. (\ref{eq:BG}) corresponds to the result of
Ref. \cite{BEF} obtained at leading nontrivial order in the chiral quark
model. We remark that the $m_\pi^2$ suppression of the matrix element is
valid only at this order, and that  terms proportional to $m_K^2$ arise at
the next order both in the $1/N_c$ and in the chiral expansion. 
Large corrections to $B_G=1$ are therefore rather plausible, and to take
them into account we will use in what follows 
$|B_G|=1-4$.
As for $B_T$, a value very close to one can be obtained for instance 
in the framework of vector meson dominance, as in \cite{RPS}. Other
estimates give very similar values (see e.g.~\cite{CIP}). 
As a conservative
range of variation for this parameter we adopt $|B_T|=0.5-2$. 
Concerning the sign of $B_T$ and $B_G$, 
the above model-dependent considerations indicate that it
is positive in both cases. We stress, however, that this 
conclusion is not based on first principles. 

Using (\ref{eq:BG}) we write the chromomagnetic
contribution to $\epp$ as\footnote{In our conventions $\Re A_0 = 3.326
  \cdot 10^{-4}$ and $F_\pi= 131$ MeV.}
\be
\Re \left({\eps' \over \eps} \right)_G = {11 \sqrt{3} \over 64 \pi} {\omega
  \over |\eps| \Re(A_0) } {m_\pi^2 m_K^2 \over F_\pi (m_s + m_d)}
{\alpha_s(m_{\tilde g}) \over m_{\tilde g}} \eta B_G  \Im \Lambda_g^-\; \; ,
\label{epspG0}
\ee
where $\eta$ contains the effect of the scaling from $m_{\tilde g}$ down to 
$m_c$ (which is the scale at which the quark masses have to be given) and
can be found in (\ref{eta}). 
Using $\alpha_s(M_Z)=0.119$ we then obtain
\be\label{epspG1}
\Re\left(\frac{\varepsilon^\prime}{\varepsilon}\right)_G
 \simeq  209 R_g \Im\Lambda_g^-~, 
\ee
where
\be\label{Rg}
R_g = \left[\frac{\alpha_s(m_\tg)}{\alpha_s(500 {\rm
      GeV})}\right]^{\frac{23}{21}} \frac{500 {\rm GeV}}{m_\tg} 
  \sqrt{R_s} B_G~.
\ee

As for the magnetic contribution to 
the  direct CP-violating component of
$K_L \rightarrow \pi^0 e^+ e^-$, we
notice that by using Eq. (\ref{eq:BT}) one can write
\be
\langle \pi^0 e^+ e^- \vert Q_\gamma^+ \vert K^0 \rangle = -
\frac{Q_d \alpha B_T}{4 \pi m_K} \langle \pi^0 e^+ e^- \vert Q_{7V}
\vert K^0 \rangle~ ,
\label{eq:memag}
\ee
where $Q_{7V(A)}=(\bar s d)_{(V-A)}(\bar e e)_{V(A)}$.
Employing the notations of \cite{BBL} and dropping for a moment
the supersymmetric contribution to $Z_{ds}$ we get
\be\label{BRKL}
  BR(K_L \to \pi^0 e^+ e^-)_{\rm dir}= 6.3 \cdot 10^{-6} \left[
\left(  \Im \lambda_t {\tilde y}_{7A}\right)^2 +
        \left( \Im \lambda_t {\tilde y}_{7V} + \Im \Lambda_g^+
        {\tilde y}_\gamma \right)^2 \right]~,
\ee
where $\frac{\alpha}{2\pi}{\tilde y}_{7V(A)}$ 
is the Wilson coefficients of $Q_{7V(A)}$ 
(the numerical values can be found in \cite{BBL}) and
${\tilde y}_\gamma$ is defined by
\bea
\Im \Lambda_g^+ {\tilde y_\gamma}  &=& 
 \frac{Q_d B_T}{\sqrt{2} G_F m_K} \Im\left[ C_\gamma^+(m_c) \right]
 \; \; , \nonumber \\
{\tilde y_\gamma}  &=& -19.3 B_T {500 \mbox{GeV} \over m_{\tilde g}}
R_{\alpha_s}^{25 \over 21} \left[ {F_0(x_{g q}) \over
    G_0(x_{g q})}
 +8 \left(1-1.13 R_{\alpha_s}^{-{2\over 21}}
    \right)   \right] \; , \label{yg1}
\eea
where $R_{\alpha_s}=\alpha_s(m_{\tilde g})/\alpha_s(500 \mbox{GeV})$.

\subsection{Supersymmetric $\epp$}
\label{sect:susyeps}
We decompose the SUSY contributions to $\epp$ as follows:
\begin{equation}
  \label{eq:eppdec}
  \Re\left( \frac{\varepsilon^\prime}{\varepsilon}
  \right)^{\mbox{\tiny{SUSY}}} = 
  \Re\left(\frac{\varepsilon^\prime}{\varepsilon}   \right)_Z
 +\Re\left(\frac{\varepsilon^\prime}{\varepsilon}   \right)_G
\end{equation}
where the first term is the contribution from the supersymmetric effective
$\sdz$ vertex and the second is the contribution of the chromomagnetic
penguin operator already discussed and given in (\ref{epspG1}).

 From \cite{BS98} we have
\begin{equation}
  \label{eq:eppz}
  \Re\left(\frac{\varepsilon^\prime}{\varepsilon} \right)_Z =  \Bigl[ 1.2 -
  R_s \vert r_Z^{(8)}\vert B_8 ^{(3/2)}\Bigr] \Im \Lambda_t \, , 
\end{equation}
where
\begin{equation}
  \label{eq:rs}
  R_s = \left[ \frac{158 {\rm MeV}}{\ms(\mc) + \md(\mc)} \right]^2
\end{equation}
and $B_8^{(3/2)}$ is the usual non-perturbative parameter
describing the hadronic matrix element of the dominant 
electroweak penguin operator. Finally $\vert 
r_Z^{(8)}\vert$ is a calculable renormalization scheme independent
 parameter in the analytic formula for
$\epp$ in \cite{bratios} which increases with $\als^{\overline {MS}}(M_Z)$
and in the range $0.116 \le \als^{\overline {MS}}(M_Z) \le 0.122$ takes
the values 
\begin{equation}
  \label{eq:rz8}
  7.1 \le  \vert r_Z^{(8)}\vert \le 8.4\,.
\end{equation}
For $R_s$ we will use the range
\begin{equation}
  \label{eq:rrs}
  1 \le R_s \le 2\,,
\end{equation}
which is compatible with the most recent lattice and QCD sum rules
calculations as reviewed in \cite{EP99}. 
Note that $R_s$ is defined as in \cite{BS98}, which differs from
\cite{EP99} where $158{\rm MeV}$ has been replaced by
$137{\rm MeV}$. Correspondingly the updated values of
$\vert r_Z^{(8)}\vert$ given in \cite{EP99} have been rescaled
appropriately.
We consider the ranges
in \r{eq:rz8} and \r{eq:rrs} as conservative. Finally we will use
as in \cite{EP99}
\begin{equation}
  \label{eq:bpars}
  0.6 \le B_8^{(3/2)} \le 1.0\, .
\end{equation}
Our treatment of all the other
parameters which enter in the SM estimate of 
$\epp$ will be explained in Section 6.

\subsection{Rare Decays}
\label{subs:rare}
Following  \cite{BS98} we have
\be
  {\rm BR}(\kppn) = {\rm BR}^+_{\SM}+1.55 \cdot 10^{-4}
\Biggl[ 2 X_0 \Re \left( \lambda_t  \Lambda_t^* \right) 
+ 2 \Delta_c \Re \Lambda_t +|\Lambda_t|^2 \Biggr]\,, 
  \label{eq:fkppn}
\ee
where ${\rm BR}^+_{\SM}$ is the Standard Model contribution given by
\be
{\rm BR}^+_{\SM}=1.55 \cdot 10^{-4}
\Biggl[ (X_0\Im\lambda_t)^2+(X_0\Re\lambda_t+ \Delta_c)^2\Biggr]\,,
\ee
where
\begin{equation}
  \label{eq:deltac}
  \Delta_c = - (2.11 \pm 0.30) \cdot 10^{-4}
\end{equation}
represents the internal charm contribution \cite{nlo1} and
$X_0=C_0-4B_0=1.52$ is the combination of penguin and box diagram functions
in (\ref{B0}) evaluated at $\mtb(\mt)=166$ GeV.
For an updated discussion about the SM estimate of the
branching ratio we refer to \cite{BB99}.

Next, following \cite{BS98} and including the contribution of the
$\gamma$-magnetic penguin to $K_L\to \pi^0e^+e^-$ we have
\begin{eqnarray}
  \label{eq:fklpn}
  {\rm BR}(\klpn) &=& {\rm BR}_{\SM}^0+  
  6.78 \cdot 10^{-4} \Bigl[ 2X_0 \Im
  \lambda_t \Im \Lambda_t+(\Im \Lambda_t)^2 \Bigr] \,, \\
  \label{eq:fkpe}
  {\rm BR}(\kpe)_{\rm dir} &=& {\rm BR}_{\SM}^{ee}+1.19 \cdot 10^{-4} 
  \biggl[  2Y_0 \Im \lambda_t \Im \Lambda_t+(\Im \Lambda_t)^2 \nn \\ 
  && + 2.13 \Im \lambda_t \bigl(0.08 \Im \Lambda_t +0.23 \Im
  \Lambda_g^+ \tilde y_\gamma \bigl) \nn \\
  && + \Bigl (0.08 \Im \Lambda_t +0.23 \Im
  \Lambda_g^+ \tilde y_\gamma \Bigr)^2 \biggr]\,, \\
  \label{eq:fkmm}
  {\rm BR}(\kmm)_{\rm SD} &=& {\rm BR}_{\SM}^{\mu\mu}
  +6.32 \cdot 10^{-3} \Bigl[
  2 \bigl( Y_0 \Re \lambda_t + \bar \Delta_c \bigl) \Re \Lambda_t \nn \\
  && +(\Re
  \Lambda_t)^2 \Bigr] \,,
\end{eqnarray}
where the Standard Model contributions are given as follows
\beqa
{\rm BR}_{\SM}^0 &=&  
  6.78 \cdot 10^{-4} \Bigl[ X_0 \Im\lambda_t \Bigr]^2 \,, \\
{\rm BR}_{\SM}^{ee} &=& 
  1.19 \cdot 10^{-4} (\Im\lambda_t)^2 
  \Bigl[Y_0^2+(1.0+0.08 C_0)^2 \Bigr]\,, \\
{\rm BR}_{\SM}^{\mu\mu} &=& 6.32 \cdot 10^{-3} \Bigl[
   Y_0 \Re \lambda_t + \bar \Delta_c\Bigr]^2 \,.
\eeqa
Here $Y_0=C_0-B_0=0.97$, $C_0=0.79$ and  
\begin{equation}
  \label{eq:deltacbar}
  \bar \Delta_c = - (6.54 \pm 0.60) \cdot 10^{-5}\,
\end{equation}
represents the charm contribution to $\kmm$
\cite{nlo1}.

Using \r{eq:fkppn}, \r{eq:fklpn} and \r{eq:fkmm} one finds the 
following useful formula \cite{BS98}
\begin{eqnarray}
  {\rm BR}(\kppn) &=& 1.55 \cdot 10^{-4} 
\Biggl[ \pm 3.97\sqrt{\kappa}\cdot 10^{-4}-3 B_0 \Re\lambda_t+
 \hat\Delta_c\Biggr]^2 \nn \\
  &&\qquad + 0.229\cdot  {\rm BR}(\klpn)~,
   \label{eq:fkppf}
\end{eqnarray}
where
\begin{equation}
\label{eq:deltahat}
 \hat\Delta_c=\Delta_c-\bar\Delta_c=- (1.46 \pm 0.30) \cdot 10^{-4}\,
\end{equation}
and $\kappa$ is defined through
\begin{equation}
\label{kappa}
  {\rm BR}(\kmm)_{\rm SD}=  \kappa \cdot 10^{-9}~.
\end{equation}
In evaluating $\hat\Delta_c$ we have included the correlation between
$\Delta_c$ and $\bar\Delta_c$ due to their simultaneous dependence
on $\Lms^{(4)}$ and $\mc$ \cite{nlo1}. The upper bound on
${\rm BR}(\kppn)$ is obtained for negative sign in (\ref{eq:fkppf})
which corresponds to $\Re \Lambda_t< C_0 |\Re \lambda_t|$
(or $\Re Z_{sd}<0$).

\section{Indirect bounds on supersymmetric contributions}
\label{sec:bounds}
\subsection{Preliminaries}
We now discuss the presently available constraints,
not directly obtained by $\epp$ or rare decays, on 
the flavour-changing mass insertions introduced in 
Section~\ref{sect:Heff}. A general model-independent
constraint on left-right mass insertions is dictated by vacuum
stability.  In particular, the requirement of avoiding charge- or
color-breaking minima or unbounded-from-below directions in the SUSY
potential implies \cite{CD} 
\beq
\label{eq:vs_bounds}
\left|\left(\delta^{D}_{LR}\right)_{12(21)}\right| 
 \lsim \frac{ \sqrt{3} m_s}{m_\tq}~, \qquad\qquad
\left|\left(\delta^{U}_{LR}\right)_{i3}\right| 
 \lsim \frac{ \sqrt{3} m_t}{m_\tq}~.
\eeq
Given the large difference between top and strange 
quark masses, the two constraints in (\ref{eq:vs_bounds}) 
are numerically very different.  However, when translated in 
bounds for the corresponding contributions to $\epp$
they look rather similar. 
Neglecting the dependence on the sparticles mass ratios, that  
is rather mild, we obtain
\beq
\label{eq:vs_bounds_2}
\left| \Lambda^{\pm}_g \right |  
\lsim  10^{-4} \left(\frac{500\rm{GeV}}{m_\tq}\right)~, \qquad\qquad
\left| \Lambda_t \right| \lsim 3 \cdot 10^{-3} 
\left(\frac{500\rm{GeV}}{m_\tq}\right)^2~,
\eeq
which leave open the possibility of large contributions to $\epp$ 
(up to $\sim 10^{-2}$) both from $\Im\Lambda^-_g$ and $\Im\Lambda_t$. 
Concerning the bound on $\Im\Lambda^+_g$, relevant to 
$K_L \to \pi^0 e^+ e^-$, we further note that
up to unlikely cancellations among $(\delta^{D}_{LR})_{12}$
and $(\delta^{D}_{LR})_{21}$ one expects
\beq
\left|\Im \Lambda^-_g \right| \sim  \left| \Im \Lambda^+_g \right| ~.
\eeq

Indirect bounds on $\Lambda^{\pm}_g$ and $\Lambda_t$
can also be obtained by $|\Delta S|=2$ amplitudes, 
barring the possibility of accidental cancellations.
In the case of $(\delta^{D}_{LR})_{12(21)}$,
the indirect constraints imposed by $\eps_K$
and $\Delta m_K$  are rather mild \cite{GGMS} and 
essentially do not affect the bound in (\ref{eq:vs_bounds}). 
In the case of $\Lambda_t$,
the constraints from $|\Delta S|=2$ amplitudes are of 
two types: those imposed by chargino box diagrams 
\cite{CI}\footnote{~We note that the chargino contribution to 
$|\Delta S|=2$ amplitudes  has been overestimated in \protect\cite{CI} 
due to a missing factor $1/4$ in the r.h.s. of Eq.~(3.4).
Moreover, though formally correct, Eq.~(3.5) of \protect\cite{CI} 
does not correspond to the expansion of $\cH_{|\Delta S|=2}$ near 
$x_{ki}=1$ (due to the missing factor $1/M^2_{\tq_k}$).
Taking into account these two corrections, we found that 
the bounds on $\tlambda_t$ in Eqs.~(3.6-7) of \protect\cite{CI} 
should be increased by a factor 3.}
and those obtained via radiative corrections, 
relating $\deulr_{23} \deulr_{13}^*$ to $(\delta^{D}_{LL})_{12}$.
It turns out that the constraints via radiative corrections 
using Renormalization Group evolution
are more severe than the ones from chargino box diagrams. We therefore
discuss the former constraints in some detail.

\subsection{Bounds on $\Lambda_t$ via Renormalization Group}
\begin{figure}[t]
\begin{center}
\leavevmode
\epsfxsize=10cm\epsfbox{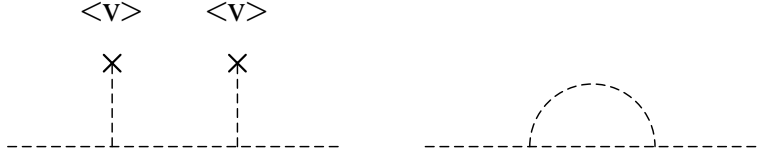}
\end{center}
\caption{\label{fig:rge} Diagrams through which the trilinear ${\bf A}^u$
  couplings may generate a $\tilde s_{L(R)} \rightarrow
  \tilde d_{L(R)}$ transition.}
\end{figure}

The presence of a large double mass-insertion of the 
type $(\tu^{d}_{L}-\ttop_{R}) \times (\ttop_{R}-\tu^s_{L})$
could have a sizable indirect effect on the 
mixing of the first two generations, that is 
strongly constrained in the down sector \cite{GGMS}.
Indeed, the trilinear  couplings ${\bf A}^{u,d}$ induce 
at one loop a flavour-changing  mass term for both left- and 
right-handed squarks, i.e. give a radiative contribution to the
off-diagonal elements of the mass matrices 
${\bf m}_Q^2$, ${\bf m}_\tu^2$ and ${\bf m}_\td^2$ \cite{BBMR}. 
The diagram which generates such an effect is depicted in
Fig. \ref{fig:rge}, together with the diagram with the double
$LR$ mass-insertion which yields the $\td_A-\ts_A$ ($A=L,R$)
transition. A naive order-of-magnitude comparison between the two
diagrams (say, at low momentum $q^2$ flowing along the squark line) would
lead one to say that the loop diagram is suppressed with 
respect to the tree one by a factor $M_S^2/ (16\pi^2 \langle v \rangle^2)
\sim 10^{-2}$, if we assume that $M_S$ is not much bigger than the
electroweak-breaking scale. However, this suppression factor, which
dominates over the finite part of the loop diagram, can be balanced by a
large logarithm arising in the divergent part of the diagram. 
In particular, in a scenario with $M_X \sim 10^{16} \gev$, the loop diagram
yields a large logarithm of the form $\ln(M_X^2/M_S^2) \sim 64$ for
$M_S \sim 10^2$ GeV, therefore compensating almost completely the
suppression factor.

To bring this discussion on more solid grounds the tools to be used are the
renormalization group equations (RGE) for the soft SUSY-breaking couplings
\cite{MV}. If we neglect all entries in the Yukawa matrices but $y_t$ and
$y_b$, the RGE for the (1,2) matrix element of ${\bf m}_Q^2$ reads as
follows ($t=\ln M_X^2/q^2$): 
\be
\label{eq:simple_rge_sol}
\frac{d ({\bf m}_Q^2)_{12}}{dt} = - {1 \over 16 \pi^2} \left({\bf A}^u 
   {{\bf A}^u}^\dagger + {\bf A}^d {{\bf A}^d}^\dagger \right)_{12} =
- {1 \over 16 \pi^2} {\bf A}^u_{13} {{\bf A}^u_{23}}^* + \ldots \; \; ,
\ee
where the ellipsis stand for terms which, according to the vacuum
stability bounds are suppressed by $(m_b/m_t)^2$ at least.
Let us now imagine for a moment that the ${\bf A}^u$ matrix elements do not 
evolve. Then we get for the radiatively generated part of $({\bf
  m}_Q^2)_{12}$:
\be\label{apbound}
({\bf m}_Q^2)^{\mbox{\tiny{rad}}}_{12}(M_S) = -\frac{ \ln(M_X^2/M_S^2)}{16
  \pi^2} {\bf A}^u_{13} {{\bf A}^u_{23}}^* \; \; ,
\ee
that, when translated into the usual $\delta$'s becomes (for $M_S=300$ GeV
and $M_X=2\cdot 10^{16}$ GeV, and $\tan \beta=5$):
\be
(\delta^{[U,D]}_{LL})^{\mbox{\tiny{rad}}}_{12} = 1.3 \cdot \deulr_{13}
\deulr^*_{23} \; \; .
\ee
(A similar expression can be obtained for the $\delta^{[U,D]}_{RR}$
couplings). 
If both the $LR$ couplings were close to the vacuum stability bounds, this
contribution would be of order one, and would violate the bounds which were 
obtained by comparison to the phenomenology of the $\Delta S=2$ transitions 
\cite{GGMS}. By reversing the argument, and assuming there is no
cancellation with the initial value of $({\bf m}_Q^2)_{12}$ at $M_X$ we
can obtain a bound on the product of the two $LR$ couplings.

In order to obtain the correct numerical value for this bound we have to
do a complete calculation and take into account also the evolution of the 
${\bf A}^u$ matrix elements. In the same approximation as above
(i.e. keeping only the $y_t$ and $y_b$ entries in the Yukawa matrices, and
neglecting all the ${\bf A}^{u,d}$ matrix elements whose vacuum-stability
bound is not proportional to $m_t$), the RGE for the ${\bf A}^u$ matrix
elements read as follows:
\be
\label{eq:Au_rge}
\frac{d {\bf A}^u_{i3}}{dt} = {1 \over 8 \pi} \left[
{16 \over 3} \alpha_3(t) + 3 \alpha_2(t) +{ 13 \over 15} \alpha_1(t) - {7
  \over 4 \pi} |y_t(t)|^2 \right] {\bf A}^u_{i3} \; \; \; \; (i\neq 3)\; \; .
\ee
The one-loop evolution of the Yukawa coupling and of the gauge coupling
constants in the MSSM is well-known, and can be found, e.g., in
\cite{MV}. The boundary conditions which we have used for these
equations are the following (for the scales $M_S=300$ GeV and $M_X=2 \cdot
10^{16}$ GeV):
\beqa
y_t(M_S) &=& 0.92 \pm 0.03 \; \; ,\nonumber \\
y_b(M_S) &=& 0.084 \; \; , \nonumber \\
\alpha_i(M_X) &=& 0.040 \pm 0.001 \; \; \; \; (i=1,\ldots, 3) \; \;
.
\eeqa
For simplicity we have evolved back from $M_X$ all three gauge couplings
from their unification value. With these boundary conditions, the solution
of the RGE equation for $({\bf m}_Q^2)_{12}(M_S)$ is the following:
\beqa
\label{eq:full_rge_sol}
({\bf m}_Q^2)_{12}(M_S) &=& ({\bf m}_Q^2)_{12}(M_X)- K  \frac{
  \ln(M_X^2/M_S^2)}{16 \pi^2} \left({\bf A}^u_{13} {{\bf
      A}^u_{23}}^*\right)(M_S) \\
K &=& (0.67 \pm 0.05) \nonumber
\; \; , 
\eeqa
where the uncertainty is mainly due to the top mass. As it is seen, the
simplified solution (\ref{apbound}) is numerically not very
different from the complete one in (\ref{eq:full_rge_sol}). It is interesting 
to note that also here the large top mass plays an important role: the Yukawa
coupling largely compensates the effect of the gauge couplings in the
evolution of the ${\bf A}^u_{i3}$ matrix elements. Neglecting the
Yukawa term in (\ref{eq:Au_rge}), the numerical coefficient $-0.67$ goes
down to $-0.34$.
Disregarding the unlikely possibility of a strong cancellation between the
two terms on the r.h.s. of (\ref{eq:full_rge_sol}) we can obtain a bound
for $\Lambda_t$ (for the numerical estimate we use again $\tan
\beta = 5$ and $M_S=300$ GeV):
\beqa
\label{eq:rge_bound}
\left| \Im\Lambda_t \right| & \leq & \frac{16 \pi^2 \sin^2
  \beta}{K \ln (M_X^2/M_S^2)} \frac{v^2 }{M_S^2 } 
   \left| H_0(x_{q\chi}) \right| {\rm min}\left\{
   \left| \Im
  (\delta^{D}_{LL} )_{12} \right|_{\mbox{\tiny{max}}},~ 
   \left| \Im
  (\delta^{U}_{LL} )_{12} \right|_{\mbox{\tiny{max}}} \right\}
  \qquad \nn \\
  & \sim & (1.2 \pm 0.1) \left| H_0(x_{q\chi}) \Im (\delta^{D}_{LL} )_{12}
  \right|_{\mbox{\tiny{max}}} \label{rgb2}
\eeqa
and analogously for the real part. 

The left-left mixing among the first two generations of
down-type squarks is strongly constrained since it appears
in gluino-mediated $|\Delta S|=2$ amplitudes \cite{GGMS}.
Since $(\delta^{D}_{LL} )_{12}$ enters quadratically in $|\Delta S|=2$
transitions, one gets the following bounds from $\Delta M_K$ and
$\ep$ respectively \cite{GGMS}:
\bea
\sqrt{\left| \Re (\delta^{D}_{LL} )^2_{12}
\right|} &\leq& 2.4\cdot 10^{-2} 
\sqrt{\left|\frac{4 f_6(1) +11 \tilde{f}_6(1)} 
{ 4 x_{gq} f_6(x_{gq}) +11 \tilde{f}_6(x_{gq})} \right|
  }~\frac{m_{\tilde q}}{300{\rm GeV}}~, \label{rgb3a} \\
\sqrt{\left| \Im (\delta^{D}_{LL} )^2_{12}
\right|} &\leq& 1.9\cdot 10^{-3} 
\sqrt{\left|\frac{4 f_6(1) +11 \tilde{f}_6(1)} 
{ 4 x_{gq} f_6(x_{gq}) +11 \tilde{f}_6(x_{gq})} \right|
  }~\frac{m_{\tilde q}}{300{\rm GeV}}~, 
\label{rgb3b}
\eea
where the functions $f_6$ and $\tilde f_6$ are defined in \cite{GGMS}.
The combination $4x f_6(x) +11 \tilde{f}_6(x)$ has a zero at $x=2.43$, 
so that close to this particular value of the gluino-squark mass 
ratio the bounds (\ref{rgb3a}-\ref{rgb3b}) become irrelevant. 
On the other hand, this value is excluded in the present scenario where
$M_X \sim 10^{16} \gev$, because the evolution of the masses via
RGE down to electroweak scales gives the condition $x_{gq} < 1.3$ for
the scalars of the first two families~\cite{RGEb}. Moreover, even if
the limits coming from gluino exchange could be evaded, the analogous
limits coming from chargino exchange, which are not much weaker, would
still hold. 

Using Eqs. (\ref{rgb2}--\ref{rgb3b}) it is possible 
to obtain bounds on $\Im\Lambda_t$ that are more stringent than the one in
Eq.~(\ref{eq:vs_bounds_2}). However, the precise size of these
constraints depends strongly on the phase of $\Lambda_t$: if the
double insertion is purely imaginary, the constraint from
$\ep$ is ineffective and $\Im\Lambda_t$ can be 
substantially larger than in
the case in which $\Re\Lambda_t$ is different from zero.

\subsection{Scanning of the SUSY parameter space and 
model-de\-pen\-dent considerations}
Taking into account the analytic bounds discussed so far, 
we will now proceed estimating the maximal allowed size
of $\Im\Lambda_t$ in terms of various SUSY parameters. 
To do so, one has to face the usual problem of scanning 
efficiently the parameter space. 
In this particular case, the phases of the relevant FCNC
parameters are crucial: as we stressed above, the
stringent constraint from $\ep$ is not effective on pure
imaginary (double) mass insertions.

To obtain an estimate of model-independent limits on SUSY
contributions, we scan randomly with uniform distribution the
parameter space corresponding to a reasonably natural determination of
$M_Z$. More precisely, we choose the relevant parameters in the
following intervals: $-300\gev < \mu < 300\gev$\footnote{We use a real
$\mu$ to avoid problems with the electric dipole moment of the
neutron.}, $100 \gev < M_2 < 250\gev$, $3 M_2 < m_{\tilde{Q}_{12}} < 5
M_2$, $M_2 < m_{\tilde{L}_{12}}<2 M_2$, $0.4\, m_{\tilde{Q}_{12}} <
m_{\tilde{t}_R}<m_{\tilde{Q}_{12}}$. Moreover we assume unification of
gaugino masses and we discard points in which
$(M_3/m_{\tilde{Q}_{12}})^2>1.3$, the charginos are lighter than
$90\gev$, the charged sleptons lighter than $80\gev$ or the gluinos
lighter than $180\gev$. The limits we get however do not
significantly depend on the details of the scanning procedure. We
focus here only on the possibility of large enhancements with respect
to the SM due to the double mass insertion contribution to $\Im
Z_{ds}$. Since the effects of single mass insertions have already been
analyzed in detail in Ref.~\cite{BRS} and have been shown to be
smaller, or at most of the same size of the SM contribution, we do not
take them into account in the present analysis.

\begin{figure}[t]
\begin{center}
\leavevmode
\epsfxsize=8cm\epsfbox{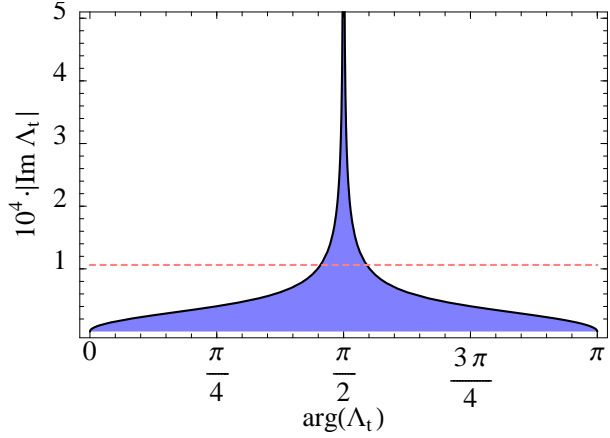}
\end{center}
\caption{Limit on $|\Im\Lambda_t|$ imposed by 
$\Delta m_K$ and $\ep$, through RGE evolution, as a function of
$\arg\Lambda_t$. The dashed line shows the SM contribution to
$\Im Z_{ds}$ for $\Im\lambda_t= 1.33 \cdot 10^{-4}$.}
\label{fig:lim}
\end{figure}

As we discussed before, the most stringent upper limits on the double
mass insertion come from $\ep$ and $\Delta m_K$ through the
RGE evolution.  To estimate the maximal possible effects, we first
choose the double mass insertion phase, then we choose the
corresponding absolute value as high as the highest limit
found scanning the parameter space. In Figure~\ref{fig:lim}, we plot
the maximal possible value of $|\Im\Lambda_t|$ as a function of
$\arg\Lambda_t$. It is evident that the stringent constraint from
$\ep$ forces $\Im\Lambda_t$ to be smaller than or of the
same order of the SM contribution to $\Im Z_{ds}$, unless
$\arg\Lambda_t$ is very close to $\pm \pi/2$. 
%
%
Therefore a large
enhancement of $\Im Z_{ds}$ with respect to the SM can only happen if
the double mass insertion is large and almost purely imaginary.  In
this particular case, combining (\ref{rgb2}) and (\ref{rgb3a}) we can
write \beq |\Im \Lambda_t| \leq 3\cdot 10^{-4} \left|
\frac{H_0(x_{q\chi})}{H_0(1)} \right| \sqrt{\left|\frac{4 f_6(1) +11
\tilde{f}_6(1)} { 4 x_{gq} f_6(x_{gq}) +11 \tilde{f}_6(x_{gq})}
\right| }~ \frac{300{\rm GeV}}{m_{\tilde q}}~.
\label{rgb4}
\eeq
As we shall discuss in the next section, this particular case can be
tested experimentally in a clear way by studying rare $K$ decays: if
for example BR$(K_L \to \pi^0 \nu \bar\nu)$ will be found to agree
with the SM expectations, then the possibility of a large
$\Im\Lambda_t$ will be ruled out.

The constraints we considered on the relevant mass insertions can be
evaded in corners of parameter space, but this holds only if an
unlikely fine-tuning is allowed. For example the limits from $\Delta
m_K$ and $\ep$ can be evaded if there is a cancellation among
the different supersymmetric contributions to them, or the limit from
RGE can be evaded if there is a cancellation between the initial value
of the insertion and the RGE contribution.

Since the insertions are pushed up to their experimental limits the
results plotted should not of course be considered as predictions but
just as maximal possible effects. Our framework is in fact general
enough to include any supersymmetric extensions of the SM with minimal
field content. This on one hand insures that we are not missing
potentially large effects.
On the other hand, one might ask whether values of $|\Im\Lambda_t|$ as
large as those ones in the shaded region of Fig.~\ref{fig:lim}
naturally arise in supersymmetric models. Unfortunately, within the
most common models this is not the case, as we will now briefly show.

Explicit models account for the strong constraints on soft
supersymmetry breaking terms in different ways. In some cases the
mechanism communicating the supersymmetry breaking guarantees that
FCNC and CP-violating processes are under control. This is the case
e.g.\ of gauge mediated supersymmetry breaking and of minimal
supergravity (SUGRA).  In other cases, further ingredients are
necessary.

In the minimal situations, a quick estimate yields
\beqa
\label{min1}
\Lambda_t & \sim & 0.3\cdot 10^{-2} \lambda_t
\frac{H_0(x_{q\chi})}{H_0(1)} \left(\frac{300\gev}{m'_S}\right)^2 \\
\label{min2}
\Lambda^\pm_g & \sim & 0.3\cdot 10^{-4} \lambda_t
\frac{G_0(x_{gq})}{G_0(1)} \left(\frac{300\gev}{m''_S}\right), \eeqa
where $M_X\sim 10^{16}$ has been used to estimate $\Lambda_g^\pm$ and
$m'_S$, $m''_S$ are dimensionful combinations of diagonal soft
parameters. Eqs.~(\ref{min1}) and~(\ref{min2}) show that $\Lambda_t$
and $\Lambda_g^\pm$ give rise to negligible effects compared to the SM
ones.

On the other hand the universality hypothesis used in minimal SUGRA
has not a compelling justification. In this and other cases in which
the mechanism generating the soft terms does not guarantee that FCNC
are under control, the potential FCNC problem must be solved by
further symmetries. From this point of view the issue of why the
scalar mass eigenstates are so degenerate or so aligned with the
corresponding fermion eigenstates is the supersymmetric version of the
issue of explaining the structure of fermion masses and mixings. If
the latter is accounted for by flavour symmetries acting on the
fermion generation indices, in a supersymmetric theory the same
symmetry acts on the corresponding scalar indices. As a consequence,
whatever is the symmetry, since the Yukawa and the corresponding soft
trilinear interactions have the same quantum numbers, the structure of
their coupling matrices is the same.  Within this class of models it
is therefore possible to show that the LR mass insertions involving
the third generation, and in turn the double insertion, are similar to
those obtained in the minimal models.  This is not what happens for
$(\delta^{D}_{LR} )_{12}$, that can be shown to be of the right order
of magnitude to generate the experimental value of
$\varepsilon'/\varepsilon$~\cite{Masiero}.  Therefore the most likely
situation, as far as the most common SUSY models are concerned, is
somewhere between the case of $\Lambda_t\simeq 0$ and $\Lambda_g\neq
0$ and the case of $\Lambda_t=\Lambda_g=0$.
We stress, however, that the flavor structure of the supersymmetry
breaking is far from having been established.  It is then worthwhile to
investigate also more exotic possibilities, like the one of a large
$\Im\Lambda_t$, as far as these are not ruled out by phenomenological
constraints.

\section{Numerical Analysis}
\label{sect:analysis}
\subsection{Strategy}
\label{subs:strat}
We are now ready to discuss magnitude and 
relations among possible supersymmetric 
contributions to $\epp$ and rare decays.
To this purpose it is useful to distinguish between
three basic scenarios:
\begin{description}
\item{{\bf Scenario A}: $[\Im\Lambda_t=0,\ \Im\Lambda_g^\pm \not=0]$.}\\
This scenario is close to what happens in most SUSY
models since, as we have seen in the previous section, 
the $\sdz$ vertex can receive sizable corrections only 
in a specific region of the parameter space. 
In this case $\epp$ can be affected only by the 
chromomagnetic operator and, as shown in Section~\ref{subs:rare}, 
among the rare modes only $K_L \to \pi^0 e^+e^-$ 
is sensitive to this SUSY contribution.
On the other hand if $\Re\Lambda_t$ is substantially 
different from zero also
$\kppn$ can be significantly affected. 
\item{{\bf Scenario B}: $[\Im\Lambda_t\not=0,\ \Im\Lambda_g^\pm=0]$.}\\
In this scenario the possibility of large corrections
to $\epp$ is not favoured from the point of view of the parameter space,
but is an interesting possibility to be investigated in a  
model-independent approach. If this is the case,
sizable effects are then expected both 
in $K_L \to \pi^0 \nu \bar{\nu}$ and $K_L \to \pi^0 e^+e^-$.
\item{{\bf Scenario C}: $[\Im\Lambda_t\not=0,\ \Im\Lambda_g^\pm\not=0]$.}\\
This represents the most general case. 
Note, however, that the requirement
of having sizable cancellations in $\epp$, between supersymmetric
contributions generated by the chromomagnetic operator and the $\sdz$
vertex, implies an additional fine-tuning with respect to scenarios A and
B.
\end{description}
We will also follow \cite{BS98} and
consider three scenarios for $\lambda_t$, which enter
Standard Model contributions and its interference with
supersymmetric contributions to rare decays and $\epp$.
Indeed there is the possibility that the value of $\lambda_t$
 is modified
by new contributions to $\eps$ and $B_{d,s}^0-\bar B_{d,s}^0$
mixings. We consider therefore three scenarios:
\bi
\item
{\bf Scenario I}: $\lambda_t$ is taken from the standard analysis of
the unitarity triangle and varied in the ranges:
\begin{equation}\label{l1}
1.05\cdot 10^{-4}\le \Im \lambda_t \le 1.61 \cdot 10^{-4}
\end{equation}
\begin{equation}\label{l2}
2.3\cdot 10^{-4}\le -\Re \lambda_t \le 3.8 \cdot 10^{-4}
\end{equation}
\item
{\bf Scenario II}: $\Im\lambda_t=0$ and $\Re\lambda_t$ is varied
            in the full range consistent with the unitarity
            of the CKM matrix:
\begin{equation}\label{l3}
1.61\cdot 10^{-4}\le -\Re \lambda_t \le 5.6 \cdot 10^{-4}
\end{equation}
 In this scenario CP violation
            comes entirely from new physics contributions.
\item
{\bf Scenario III}: $\lambda_t$ is varied
            in the full range consistent with the unitarity
            of the CKM matrix:
\begin{equation}
-1.73\cdot 10^{-4}\le \Im \lambda_t \le 1.73 \cdot 10^{-4}
\end{equation}
 This means in particular that
            $\Im\lambda_t$ can be negative.
\end{itemize}
We would like to emphasize that the scenarios II and in particular III 
are very unlikely
and are presented here only for completeness. 
We stress that if one uses the Standard Model expressions for $B^0-\bar
B^0$ mixings, $\ep$ and $\sin 2\beta$ one gets results for the CKM matrix
which are compatible with the $|V_{ub}/V_{cb}|$ constraint, which is
insensitive to new physics. In view of the coherence of the Standard Model
picture, corrections to the processes in question so large as to make
$\Im\lambda_t$ negative, or $\Re\lambda_t$ way outside the range in
Eq. (\ref{l2}) look rather improbable.
We believe that if the new physics has an impact on the
usual determination of $\lambda_t$, the most likely situation is
between scenarios I and II.

\subsection{$\epsp/\eps$}
We shall now proceed extracting ranges for the effective 
SUSY couplings from the 
experimental data on $\epp$ in the basic  scenarios A-C defined above.
These will then be used to estimate the branching ratios 
of the rare decay modes.

Assuming that the SM contribution to $\Re(\epsp/\eps)$ is around its central
value, as given in \cite{EP99}, and therefore much smaller than the
experimental result, there is a lot of room for SUSY to contribute to this
quantity. 
Detailed bounds on $\Re(\epsp/\eps)^{^{\rm SUSY}}$ depend 
on the various parameters entering  $\Re(\epsp/\eps)^{^{\rm SM}}$,
as well as on the experimental result in (\ref{ga}),
however, as a simplified starting point for our discussion,
we assume at first
\beq  
\Re \left(\frac{\epsp}{\eps}\right)^{\mbox{\tiny SUSY}} =
  2 \cdot 10^{-3}~.
\label{eps_susy}
\eeq
This value has to be taken only as a reference figure:
it could be interpreted either as the difference 
between the experimental result and the SM contribution 
or as the true value of $\Re(\epsp/\eps)$ in the limit 
of a real CKM matrix.
\begin{figure}[t]
\begin{center}
\leavevmode
\epsfxsize=10cm\epsfbox{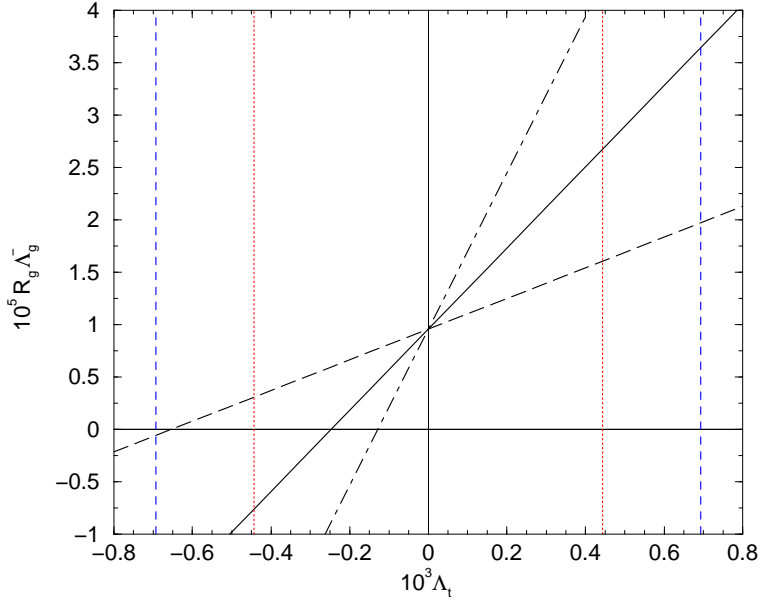}
\end{center}
\caption{\label{fig:epp} 
  Linear relation between $\Lambda_g^-$ and
  $\Lambda_t$ for 
  $\Re (\epsp/\eps)^{^{\rm SUSY}}=2\cdot~10^{-3}$. 
  The solid line is for 
  $\{ B_8^{(3/2)}, R_s, r_Z^{(8)} \}=\{0.8,1.5,7.8 \}$,
  the dot-dashed for $\{ B_8^{(3/2)}, R_s, r_Z^{(8)} \}=\{1.0,2.0,8.4 \}$ 
  and the dashed for  $\{ B_8^{(3/2)}, R_s, r_Z^{(8)}
  \}=\{0.6,1.0,7.1 \}$. The vertical lines show the RGE bound 
  (\protect\ref{rgb4}) for $m_\tq=300$~GeV and
  $\{ x_{q\chi}, x_{gq} \}=\{3,1 \}$ (dotted)
  or  $\{ x_{q\chi}, x_{gq} \}=\{9,1.3 \}$ (dashed).}
\end{figure}

Since our formula for the SUSY contribution Eq. (\ref{eq:eppdec}) contains
only two free parameters, $\Im \Lambda_t$ and $\Im \Lambda^-_g$,
Eq.~(\ref{eps_susy}) defines a straight line in the $(\Im \Lambda_t,\Im
\Lambda^-_g)$ plane, which represents the general solution within scenario
C.  This is shown in Fig.~\ref{fig:epp} for three different sets of $\{
B_8^{(3/2)}, R_s, r_Z^{(8)} \}$. Decreasing the reference value in
(\ref{eps_susy}) corresponds to a translation of the straight lines toward
the origin; the intercepts of the lines with vertical and horizontal axes
define the solutions within scenarios A and B, respectively.

As it can be noticed, if $\Lambda^-_g=0$, then $\Im \Lambda_t$ must be
negative, i.e. the SUSY contribution to the $\sdz$ vertex must be
opposite to the SM one in order to produce a positive contribution to
$\epp$.  The minimum value of $|\Im \Lambda_t|$ with $\Lambda^-_g=0$ is
found for the maximum values of $B_8^{(3/2)}$, $R_s$ and $r_Z^{(8)}$. In
this case SUSY and SM contributions to the $\sdz$ vertex cancel almost
completely and the experimental value for $\epp$ is roughly reproduced by
QCD penguin contributions. On the other hand, the maximum allowed value of
$|\Im \Lambda_t|$ with $\Lambda^-_g=0$ is found for the minimum set of $\{
B_8^{(3/2)}, R_s, r_Z^{(8)} \}$.  In this case the $\sdz$ vertex has
an opposite sign with respect to the SM case and is largely dominated by
SUSY contributions ($|Z_{ds}/Z^{\rm SM}_{ds}| \gsim 6)$. This solution is
still allowed by the RGE constraint (\ref{rgb4}), provided the sparticle
masses are not too high.  The situation of course changes if one allows
also $\Im \Lambda^-_g$ to be different from zero. In particular, for large
($\gsim 10^{-5}$) and positive values of $R_g \Im \Lambda^-_g$ a positive
$\Im\Lambda_t$ is needed in order to avoid too large effects in $\epp$.

In the limit where the standard determination of the CKM matrix is valid, a
quantitative estimate of the ranges for $\Im \Lambda^-_g$ and
$\Im\Lambda_t$, within scenarios A and B, can be obtained by subtracting
the SM contribution from the experimental value in (\ref{ga}).  Following
\cite{BS98}, we parametrize the SM result for $\epp$ using the approximate
formula
\beq
\Re \left(\frac{\varepsilon^\prime}{\varepsilon} \right)^{\mbox{\tiny SM}}
  = \Im \lambda_t \biggl[ -1.4 + R_s \Bigl[1.1 \vert r_Z^{(8)}\vert
  B_6^{(1/2)} + (1.0 - 0.67 \vert r_Z^{(8)}\vert )
  B_8^{(3/2)}\Bigr]\biggr]\, 
\label{eps_SM}
\eeq
with \cite{EP99}
\beq
 \Im \lambda_t = (1.33 \pm 0.14) \cdot 10^{-4}~.
\eeq
Varying $\Im \lambda_t$ and the experimental value 
(\ref{ga}) within $2\sigma$ intervals, choosing 
$B_8^{(3/2)}$, $R_s$ and $r_Z^{(8)}$ as discussed in
Section~\ref{sect:susyeps}
and, finally, assuming $0.7 \leq B^{(1/2)}_6 \leq 1.3$,
we find:
\beqa
-15.5 \cdot 10^{-4} \leq& 
\Re \left(\frac{\epsp}{\eps}\right)^{\mbox{\tiny SUSY}} &\leq
 30.1 \cdot 10^{-4}\; , \label{epe_SUSY_2}\\
-9.3 \cdot 10^{-4} \leq&  \Im \Lambda_t &\leq
   1.7 \cdot 10^{-4} \qquad (\Im \Lambda_g^- =0)\; ,
\label{Lt_eps} \\
-0.7 \cdot 10^{-5} \leq& R_g \Im \Lambda_g^- &\leq
   1.4 \cdot 10^{-5} \qquad (\Im \Lambda_t =0) \; .
\label{LG}
\eeqa

It is interesting to note that the range of $\Im \Lambda_g^-$ 
is well within the bound (\ref{eq:vs_bounds}), therefore
$\epp$ provides the most stringent bound on $|\Im \Lambda_g^-|$
within scenario A.
Similarly, $\epp$ provides the most stringent 
model-independent upper bound 
on $\Im \Lambda_t$ within scenario B. 
On the other hand, the lower bound on $\Im \Lambda_t$
imposed by $\epp$ is weaker than the bound 
(\ref{rgb4}) for $m_\tq \gsim 200$~ GeV and $x_{gq}<1.3$.

To show the possible improvement due to more precise measurement
of $\epp$ we show how (\ref{epe_SUSY_2})--(\ref{LG})
are modified if we fix $\Re(\epp)_{\rm exp}=20\cdot 10^{-4}$.
We find
\beqa
-7.5 \cdot 10^{-4} \leq& 
\Re \left(\frac{\epsp}{\eps}\right)^{\mbox{\tiny SUSY}} &\leq
 19.7 \cdot 10^{-4}\; , \label{epe1_SUSY_2}\\
-5.9 \cdot 10^{-4} \leq&  \Im \Lambda_t &\leq
   0.8 \cdot 10^{-4} \qquad (\Im \Lambda_g^- =0)\; ,
\label{Lt1_eps} \\
-0.4 \cdot 10^{-5} \leq& R_g \Im \Lambda_g^- &\leq
   0.9 \cdot 10^{-5} \qquad (\Im \Lambda_t =0) \; .
\eeqa

\subsection{Rare Decays}
The rare decays $K_L \to \pi^0 \nu\bar{\nu}$ and $K_L \to \pi^0 e^+e^-$
provide in principle a powerful tool to clearly establish possible SUSY
contributions in $CP$-violating $|\Delta S|=1$ amplitudes, and also to
distinguish among the three scenarios introduced in
Section~\ref{subs:strat}.
\subsubsection{Scenario A}
Within scenario A only $K_L \to \pi^0 e^+e^-$ among these two
modes is affected by SUSY corrections. 
Setting $R_{\alpha_s}=1$ in (\ref{yg1}) we can write 
\beq
\Im \Lambda_g^+ {\tilde y_\gamma}  
= 35.5 ~ R_g \Im \Lambda_g^- 
~\left[ \frac{\Im \Lambda_g^+}{\Im \Lambda_g^-}\right]
~\left[ \frac{ B_T }{ B_G \sqrt{R_S} } \right]~,
\label{yg2}
\eeq
where the numerical coefficient has been obtained for $x_{gq}=1$
and can increase at most to 37.0 if we impose $x_{gq} < 1.3$.
Assuming $R_g \Im \Lambda_g^- = 10^{-5}$,
as obtained from Fig.~\ref{fig:epp}, and 
fixing to unit the two ratios among square brackets in 
(\ref{yg2}), we obtain $\Im \Lambda_g^+ {\tilde y_\gamma} 
= 3.5 \cdot 10^{-4}$. 
Using this figure in (\ref{eq:fkpe}) we find that the 
additional contribution to ${\rm BR}(\kpe)_{\rm dir}$ 
is positive and ranges 
between 3 and 4 in units of $10^{-12}$, depending on the 
value of $\Im\lambda_t$. This effect, which represents the 
typical size of the SUSY contribution to $K_L \to \pi^0 e^+e^-$ 
expected within scenario A, is certainly difficult to be 
observed. However, we stress that this conclusion 
depends strongly on the assumptions made for 
$\Im \Lambda_g^+/\Im \Lambda_g^-$ and $B_T/(B_G \sqrt{R_S})$. 

According to the ranges of $B_T$, $B_G$ and $R_s$ 
discussed in Section~\ref{sec:BF}, we expect
\beq
0.09\leq \frac{B_T}{B_G \sqrt{R_S}} \leq 2 \; . 
\eeq
On the other hand, it is more difficult to 
estimate $\Im \Lambda_g^+/\Im \Lambda_g^-$
without specific assumptions on the 
SUSY soft-breaking terms. In minimal models 
it is natural to assume $(\delta^{D}_{LR})_{12} \gg 
(\delta^{D}_{LR})_{21}$, that implies 
\beq
\frac{\Im \Lambda_g^+ }{ \Im \Lambda_g^-} \simeq -1\; ,
\label{IpIm}
\eeq
but we cannot exclude sizable deviations from 
this figure in generic scenarios.

In Table~\ref{tab:KLee} we report the 
upper bounds on ${\rm BR}(\kpe)_{\rm dir}$, 
for different values of the two ratios.
To this end we have used the expressions
for $\epp$ and ${\rm BR}(\kpe)_{\rm dir}$
given in Section 4
with the Standard Model contribution for $\epp$ given in
(\ref{eps_SM}). 
Scanning the parameters
$B_8^{(3/2)}$, $B^{(1/2)}_6$, $R_s$ and $r_Z^{(8)}$ as discussed in
Section~\ref{sect:susyeps} and 6.2, 
varying $\Im \lambda_t$ according to (\ref{l1}) (scenario I),
we find the results in the third and fourth column
which correspond to two choices of $\epp$. As it can be noticed, 
results in the ball park of $10^{-11}$ cannot be excluded 
even under the assumption (\ref{IpIm}).

\begin{table}[p]
\begin{center}
\begin{tabular}{|c|c||c|c|} 
\hline
  $\Im \Lambda_g^+/\Im \Lambda_g^-$    &  
  $B_T/(B_G \sqrt{R_S})$  &   ${\rm BR}(\kpe)_{\rm dir}$
 & ${\rm BR}(\kpe)_{\rm dir}$
 \\ \hline\hline
  -1  & 1.0 &$ 9.4 \cdot 10^{-12}$ & $7.8 \cdot 10^{-12}$ \\ \hline
  -1  & 0.5 &$ 7.8 \cdot 10^{-12}$ & $ 7.0 \cdot 10^{-12}$ \\ \hline
  -1  & 1.5 & $ 1.1 \cdot 10^{-11}$ & $8.5 \cdot 10^{-12}$ \\ \hline
  -2  & 1.5 & $ 1.8 \cdot 10^{-11}$ & $1.1 \cdot 10^{-11}$ \\ \hline
  1  & 1.0 & $1.3 \cdot 10^{-11}$ & $1.0 \cdot 10^{-11}$ \\ \hline
  1  & 0.5  & $9.3 \cdot 10^{-12}$ & $8.2 \cdot 10^{-12}$ \\ \hline
  1  & 1.5 & $1.8 \cdot 10^{-11}$ & $1.3 \cdot 10^{-11}$ \\ \hline
  2  & 1.5 & $3.7 \cdot 10^{-11}$ & $2.3 \cdot 10^{-11}$ \\ \hline
\end{tabular}
\caption{Upper bounds on 
${\rm BR}(\kpe)_{\rm dir}$ within scenario A, 
for different values of $\Im \Lambda_g^+/\Im \Lambda_g^-$ and
$B_T/(B_G \sqrt{R_S})$ consistent with 
$12\le 10^4\Re(\epp)\le 30.4$ (third column)
and $\Re(\epp)=20.0 \cdot 10^{-4}$ (fourth column). 
The bounds are obtained setting $x_{gq}=1.3$ in order 
to maximize the numerical coefficient in (\ref{yg2}).
To maximize the interference of SM and SUSY amplitudes, 
$R_g \Im \Lambda_g^- $ is chosen as the maximum (minimum) 
value allowed by $\epp$ for 
positive (negative) $\Im \Lambda_g^+/\Im \Lambda_g^-$.}
\label{tab:KLee}
\end{center}
\end{table}

\begin{table}[p]
\begin{center}
\begin{tabular}{|c||c|c|} 
\hline
  $\Im \Lambda_g^+/\Im \Lambda_g^-$   
  & ${\rm BR}(\kpe)_{\rm dir}$ (II) 
  & ${\rm BR}(\kpe)_{\rm dir}$ (III) \\ \hline\hline
  -1  &  1.8 (0.8) $\cdot 10^{-12}$  &  2.5 (2.1) $\cdot 10^{-11}$ \\ \hline
  -2  &  7.3 (3.2) $\cdot 10^{-12}$  &  5.7 (4.5) $\cdot 10^{-11}$ \\ \hline
   1  &  1.8 (0.8) $\cdot 10^{-12}$ &   1.5 (1.2) $\cdot 10^{-11}$ \\ \hline
   2  &  7.3 (3.2) $\cdot 10^{-12}$  &  2.5 (1.7) $\cdot 10^{-11}$ \\ \hline
\end{tabular}
\caption{Upper bounds on 
${\rm BR}(\kpe)_{\rm dir}$ within scenario A
for $\Im\lambda_t=0$ (II) and $|\Im\lambda_t|< 1.73\cdot 10^{-4}$ (III).
The bounds are obtained setting 
$B_T/(B_G \sqrt{R_S})=1$ and imposing 
$\Re(\epp)\le 30.4 (20.0) \cdot 10^{-4}$.}
\label{tab:KLee2}
\end{center}
\end{table}

The dependence of ${\rm BR}(\kpe)_{\rm dir}$
%
%
on the value of $\Im\lambda_t$ 
is shown in Table~\ref{tab:KLee2}.
If the CKM matrix is real and 
$|(\Im \Lambda_g^+/\Im \Lambda_g^-) B_T/(B_G \sqrt{R_S})| \gsim 1$,
we find ${\rm BR}(\kpe)_{\rm dir} \sim {\rm few}\times 10^{-12}$,
similarly to the SM case. On the other hand 
values substantially larger than $10^{-11}$ are obtained 
within scenario III. Note, however, that 
the large results quoted for  
$\Im \Lambda_g^+/\Im \Lambda_g^- <0 $ 
are very unlikely, since are obtained
for the maximum negative value of $\Im\lambda_t$.

Concerning $\klpn$, its branching ratio in scenario A stays close to
the Standard Model value provided the usual determination of
$\Im\lambda_t$ is not substantially decreased through supersymmetric
contributions to $\ep$. Because of the unitarity of the CKM
matrix $\Im\lambda_t$ can only be marginally increased over its
SM value. On the other hand if $\Im\lambda_t=0$ a clear signature for
scenario A would be a  vanishingly small ${\rm BR}(\klpn)$ 
($\lsim 10^{-14}$ \cite{BI}).

The case of $\kppn$ is different as it is dominantly governed by 
$\Re\lambda_t$ and $\Re\Lambda_t$. The upper bound on
${\rm BR}(\kppn)$ can be obtained by using equation
(\ref{eq:fkppf}) together with the bound 
\cite{CI,dambrosio,pich,BS98}
\begin{equation}\label{KLSD}
{\rm BR}(\kmm)_{\rm SD}\le 2.8 \cdot 10^{-9}
\end{equation}
i.e. $\kappa=2.8$. 
Choosing then 
$(-\Re\lambda_t)_{\rm max}=3.8\cdot 10^{-4}$ (scenario I for $\lambda_t)$,
as obtained in the Standard
Model, or  $(-\Re\lambda_t)_{\rm max}=5.6\cdot 10^{-4}$ (scenarios II and
III), we find respectively
\begin{equation}\label{b1}
{\rm BR}(\kppn)\le 1.70\cdot 10^{-10}+0.229 {\rm BR}(\klpn)~,
\end{equation}
\begin{equation}\label{b2}
{\rm BR}(\kppn)\le 2.03\cdot 10^{-10}+0.229 {\rm BR}(\klpn)
\end{equation}
As the second terms on the r.h.s of these bounds are very small
in this scenario we find ${\rm BR}(\kppn)\le 1.7\cdot 10^{-10}$
and ${\rm BR}(\kppn)\le 2.1\cdot 10^{-10}$. These results are
also obtained if $\Re\Lambda_t$ is varied in the full range
consistent with the bound (\ref{KLSD}) and with the RGE
constraint (\ref{RG1}) with $\Im\Lambda_t=0$. Evidently as
(\ref{b1}) and (\ref{b2}) have been obtained without the
constraint (\ref{RG1}), what matters in this scenario is
(\ref{KLSD}).

\subsubsection{ Scenario B}
Being strongly sensitive to $\Im\Lambda_t$ and insensitive to
$\Im\Lambda_g^\pm$, $\klpn$ represents the golden mode to identify
scenarios B and C.  We first discuss scenario B which corresponds to the
case analyzed in \cite{BS98}. This time, however, the effective $\sdz$
vertex is additionally constrained by the renormalization group analysis of
Section 5.
  
\begin{figure}[t]
\begin{center}
\leavevmode
\epsfxsize=15cm\epsfbox{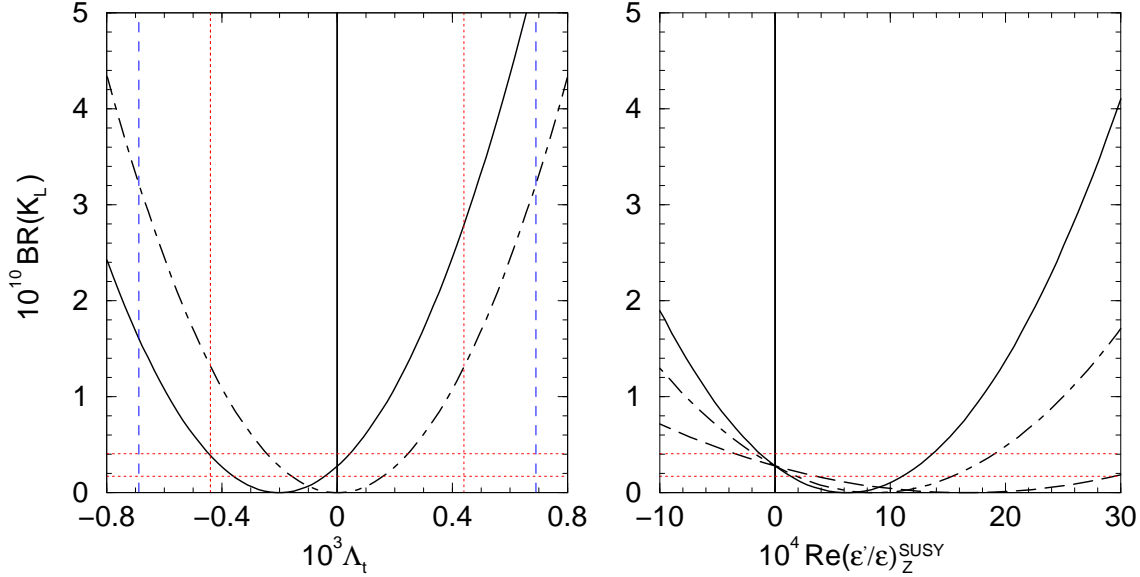}
\end{center}
\caption{\label{fig:KLvsEE} ${\rm BR}(K_L \rightarrow\pi^0 \nu \bar\nu $) 
  as a function of $\Im \Lambda_t$ (left) or as a 
  function of  $(\epp)_Z^{^{\rm SUSY}}$ (right). In the left 
  plot the solid (dot-dashed) parabola is for 
  $\Im \lambda_t= 1.33 \cdot 10^{-4}$ (0) and
  the vertical lines show the RGE bounds
  as in Fig.~\protect\ref{fig:epp}.
  In the right plot the three parabola are for
  $\Im \lambda_t= 1.33 \cdot 10^{-4}$ and 
  $\{ B_8^{(3/2)}, R_s, r_Z^{(8)} \}=\{0.6,1.0,7.1 \}$ (solid),
  $\{ B_8^{(3/2)}, R_s, r_Z^{(8)} \}=\{0.7,1.0,7.8 \}$ (dot-dashed)
  or  $\{ B_8^{(3/2)}, R_s, r_Z^{(8)}\}=\{0.8,1.5,7.8 \}$ (dashed).
  In both cases the horizontal lines denote the SM range 
  of ${\rm BR}(K_L \rightarrow\pi^0 \nu \bar\nu $) 
  for $1.05< 10^4 \Im \lambda_t < 1.61$.  
}  
\end{figure}

The dependence of ${\rm BR}(\klpn)$ on $\Im\Lambda_t$ is shown in the left
plot of Fig.~\ref{fig:KLvsEE}. As can be noticed, large enhancements
with respect to the SM case are possible, but on the other hand we cannot
exclude a destructive interference among SUSY and SM contributions leading
to strong suppression of ${\rm BR}(\klpn)$.

If the standard determination of $\Im \lambda_t$ is valid, 
Eq.~(\ref{eq:fklpn}) implies that 
${\rm BR}(K_L \rightarrow \pi^0 \nu \bar\nu)$ can be 
enhanced with respect to the SM case only if
\beq
 \Im \Lambda_t < - 2 X_0 \Im  \lambda_t 
\qquad {\rm or} \qquad  \Im \Lambda_t > 0~. 
\eeq
The second possibility is excluded within scenario B
if we require a positive SUSY contribution to $\epp$.
This is clearly shown by the second plot
in Fig.~\ref{fig:KLvsEE}, which illustrates the 
relation between ${\rm BR}(K_L \rightarrow \pi^0 \nu \bar\nu)$
and the SUSY contribution to $\epp$
within scenario B, assuming the standard 
determination of $\Im \lambda_t$.
In this case large enhancements of 
${\rm BR}(K_L \rightarrow \pi^0 \nu \bar\nu)$
are possible, but only if $R_s$ and $B_8$ 
are close to their minimum values. 
On the other hand, if $R_s$ and/or $B_8$ 
are large, then ${\rm BR}(K_L \rightarrow \pi^0 \nu \bar\nu)$
is more likely to be suppressed rather than enhanced with 
respect to the SM case.

In order to be more quantitative 
we consider the three scenarios for $\lambda_t$ defined at the
beginning of this section.
Next, as discussed in Section 6.2, $\Im\Lambda_t$ can be best
bounded by $\epp$ and the renormalization group analysis 
of Section 5. 
$\Re\Lambda_t$ can be bounded by the present
information on the short distance contribution to $\kmm$
and also by the RG analysis of Section 5, as we will state more
explicitly below. These
bounds imply a bound on ${\rm BR}(\kppn)$. Since ${\rm BR}(\kppn)$
depends on both $\Re\Lambda_t$ and $\Im\Lambda_t$ also the bound
on $\Im\Lambda_t$  matters in cases where it is substantially
larger than the Standard Model contribution to $\Im Z_{ds}$.

The branching ratios ${\rm BR}(\klpn)$ and ${\rm BR}(\kpe)_{\rm dir}$ are
dominated by $(\Im Z_{sd})^2$.  Yet, the outcome of this analysis depends
sensitively on the sign of $\Im Z_{sd}$. Indeed, $\Im Z_{sd}>0$ results in
the suppression of $\epp$ and since in the Standard Model the value for
$\epp$ is generally below the data, substantial enhancements of $\Im
Z_{sd}$ with $\Im Z_{sd}>0$ are not possible. The situation changes if new
physics reverses the sign of $\Im Z_{sd}$ so that it becomes negative.
Then the upper bound on $-\Im Z_{sd}$ is governed by the upper bound on
$\epp$ and with suitable choice of hadronic parameters and $\Im\lambda_t$
(in particular in scenario III) large enhancements of $-\Im Z_{sd}$ and of
rare decay branching ratios are in principle possible. The largest
branching ratios are found when the neutral meson mixing is dominated by
new physics contributions which force $\Im\lambda_t$ to be as negative as
possible within the unitarity of the CKM matrix.  As we argued above, this
possibility is quite remote. However, if this situation could be realized
in some exotic model, then the branching ratios in question could be very
high as demonstrated in \cite{BS98}.

In this context it is interesting to observe that in the case of
supersymmetry such large enhancements of $-\Im Z_{sd}$ while allowed by
$\epp$ are ruled out by the renormalization group bound on $\Im\Lambda_t$
considered in Section 5. As we will see in a moment the imposition of the
bound (see Fig.~\ref{fig:lim})
\begin{equation}\label{RG}
|\Im\Lambda_t|\le 5.0\cdot 10^{-4}
\end{equation}
has in the case of a negative $\Im\Lambda_t$ a very large impact on the
analysis in \cite{BS98} suppressing considerably the upper bounds on rare
decays obtained there.

\begin{table}[p]
\begin{center}
  \begin{tabular}{|c||l|c|c|c|c|}
  \hline
  $ 10^{4}\Re(\epp)_{\rm min} $ & \multicolumn{1}{c}{Scenario for $\lambda_t$:}
  & \multicolumn{1}{c}{I} & \multicolumn{1}{c}{II} & III  & SM  \\ \hline
  & $10^{10}~ {\rm BR}(\klpn)$ & $1.2~(0.6)$ &$-$ & $1.4~(0.8)$ & 0.4 \\  
  12.0 & $10^{11}~ {\rm BR}(\kpe)_{\rm dir}$  & $1.7~(0.9)$ &$-$ 
       & $2.1~(1.1)$ & 0.7\\  
  & $10^{10}~ {\rm BR}(\kppn)^*$ & $2.0~(1.8)$ &$-$ & $2.4~(2.2)$ & 1.1 \\ 
  & $10^{10}~ {\rm BR}(\kppn)$   & $1.7~(1.7)$ &$-$ & $2.1~(1.9)$ & 1.1 \\ 
  \hline 
  & $10^{10}~ {\rm BR}(\klpn)$ & $0.7~(0.4)$ &$-$ & $0.9~(0.5)$ & 0.4 \\  
  20.0 & $10^{11}~{\rm BR}(\kpe)_{\rm dir}$  & $1.1~(0.7)$ &$-$ 
       & $1.3~(0.8)$ & 0.7  \\ 
  & $10^{10}~{\rm BR}(\kppn)^*$ & $1.9~(1.8)$ &$-$ & $2.2~(2.2)$ & 1.1  \\
  & $10^{10}~{\rm BR}(\kppn)$   & $1.7~(1.7)$ &$-$ & $2.0~(1.9)$ & 1.1  \\ 
 \hline 
  \end{tabular}
\end{center}
\caption{Upper bounds for the branching ratios 
   of the rare decays $\klpn$, $\kpe$ and
   $\kppn$ in the case $\Im\Lambda_t>0$, $\Im\Lambda^\pm_g=0$.
   The results have been  obtained 
   in various scenarios for $\lambda_t$ by imposing 
   $\Re(\epp)\ge 12.0  \cdot 10^{-4}$ 
   or $\Re(\epp)\ge 20.0  \cdot 10^{-4}$, 
   with $B_8^{(3/2)}=0.6 (1.0)$. The $^*$ means that the ${\rm BR}(\kppn)$
   has been calculated using the bounds (\protect{\ref{b1}}) and
   (\protect{\ref{b2}}). Otherwise, the more stringent bound due to RGE,
   Eq. (\protect{\ref{RG1}}), has been used.
  \label{tab:rarepo1}}
\end{table}

\begin{table}[p]
\begin{center}
  \begin{tabular}{|c||l|c|c|c|c|}
  \hline
  $ 10^{4}\Re(\epp)_{\rm max} $ & \multicolumn{1}{c}{Scenario for $\lambda_t$:}
  & \multicolumn{1}{c}{I} & \multicolumn{1}{c}{II} & III  & SM  \\ \hline 
  & $10^{10}~ {\rm BR}(\klpn)$ &  $0.8~(0.8)$ &$1.7~(1.7)$ & 
    $4.0~(4.0)$ & 0.4 \\  
  30.4 & $10^{11}~ {\rm BR}(\kpe)_{\rm dir}$ & $2.0~(2.0)$ &$3.0~(3.0)$ & 
    $5.9~(5.9)$ & 0.7  \\  
  & $10^{10}~ {\rm BR}(\kppn)^*$ & $1.9~(1.9)$ &$2.4~(2.4)$ & 
    $2.9~(2.9)$ & 1.1  \\ 
  & $10^{10}~ {\rm BR}(\kppn)$ & $1.7~(1.7)$ &$2.1~(2.1)$ & 
    $2.7~(2.7)$ & 1.1  \\ 
 \hline 
  & $10^{10}~ {\rm BR}(\klpn)$ & $0.8~(0.4)$ & $1.7~(0.8)$ & 
    $4.0~(3.8)$ & 0.4 \\  
  20.0 & $10^{11}~ {\rm BR}(\kpe)_{\rm dir}$  & $2.0~(0.7)$ &$3.0~(1.4)$ & 
    $5.9~(5.7)$ & 0.7  \\  
  & $10^{10}~ {\rm BR}(\kppn)^*$ & $1.9~(1.8)$ &$2.4~(2.2)$ & 
    $2.9~(2.9)$ & 1.1  \\ 
  & $10^{10}~ {\rm BR}(\kppn)$ & $1.7~(1.7)$ &$2.1~(1.9)$ & 
    $2.7~(2.6)$ & 1.1  \\
 \hline 
  \end{tabular}
\end{center}
\caption{Upper bounds for the branching ratios 
   of the rare decays $\klpn$, $\kpe$ and
   $\kppn$ in the case $\Im\Lambda_t<0$, $\Im\Lambda^\pm_g=0$.
   The results have been  obtained 
   in various scenarios for $\lambda_t$ by imposing 
   $\Re(\epp)\le 30.4  \cdot 10^{-4}$ 
   or $\Re(\epp)\le 20.0  \cdot 10^{-4}$, 
   with $B_8^{(3/2)}=0.6 (1.0)$. For an explanation of the $^*$ see caption 
   of Table \protect{\ref{tab:rarepo1}}.
  \label{tab:rarepo3}}
\end{table}

\begin{table}[t]
\begin{center}
  \begin{tabular}{|l|c|c|c|c|}
  \hline 
  \multicolumn{1}{|c}{Scenario for $\lambda_t$:}
  & \multicolumn{1}{c}{I} & \multicolumn{1}{c}{II} & III  & SM  \\ \hline 
  $10^{10}~{\rm BR}(\klpn)$ & $3.9$ &$6.5$ & $17.6$ & 0.4 \\ \hline 
  $10^{11}~{\rm BR}(\kpe)_{\rm dir}$  
         & $7.9$ &$11.5$ & $28.0$ & 0.7  \\ \hline 
  $10^{10}~{\rm BR}(\kppn)$ & $2.6$ &$3.5$ & $6.1$ & 1.1  \\ \hline 
  \end{tabular}
\end{center}
  \caption{Upper bounds for the branching ratios 
   of the rare decays $\klpn$, $\kpe$ and
   $\kppn$ in scenario B, without imposing 
   the RGE constraint (\ref{RG}) and 
   using $B_8^{(3/2)}=0.6 $.
   \label{tab:rarepo5}}
\end{table}

In Table~\ref{tab:rarepo1}   
we show the upper bounds on rare decays for
$\Im \Lambda_t>0$ for three scenarios of $\Im\lambda_t$ in question
and two different lower bounds on $\epp$.
To this
end all parameters relevant for $\epp$ have been scanned
in the ranges used in scenario A except that $\Im \Lambda^\pm_g$
have been set to zero.
In Table~\ref{tab:rarepo3} 
the case $\Im\Lambda_t<0$ for two different upper bounds on $\epp$
is considered.
In the last column we always give the upper bounds obtained
in the Standard Model.

The inspection of Table~\ref{tab:rarepo1} shows that only moderate
enhancements of branching ratios are allowed by $\epp$ if
$\Im\Lambda_t>0$. Moreover the case $\Im \lambda_t=0$ is excluded by the
positive value of $\epp$. If $\Im \Lambda_t<0$, substantial enhancements of
${\rm BR}(\klpn)$ and ${\rm BR}(\kpe)_{\rm dir}$ are possible as seen in
Table~\ref{tab:rarepo3}.  In particular in scenario III both branching
ratios can be enhanced by one order of magnitude over Standard Model
expectations. On the other hand the imposition of the the RGE bound
(\ref{RG}) plays an important role in this analysis.  In
Table~\ref{tab:rarepo5} we show what one would find instead of
Table~\ref{tab:rarepo3}, for $\Re(\epp)_{\rm max}=30.0\cdot 10^{-4}$, if
the bound (\ref{RG}) had not been imposed.  Table~\ref{tab:rarepo5}
corresponds to the analysis in \cite{BS98} and shows very clearly that
without the bound (\ref{RG}) very large enhancements of branching ratios in
question are possible.  One should note the strong sensitivity of the
results to the choice of $B_8^{(3/2)}$ in Tables~\ref{tab:rarepo1} and
\ref{tab:rarepo5}, where the bounds are governed by $\epp$. On the other
hand this sensitivity is absent in Table~\ref{tab:rarepo3} for
$\Re(\epp)_{\rm max}=30.0\cdot 10^{-4}$ and in scenario III for
$\Re(\epp)_{\rm max}=20.0\cdot 10^{-4}$, where the bounds on $\klpn$ and
$\kpe$ are governed by the renormalization group bound (\ref{RG}).

Next we should make a few remarks on $\kppn$. The bounds on ${\rm
BR}(\kppn)$ denoted by ``*" in Tables~\ref{tab:rarepo1} and
\ref{tab:rarepo3} have been obtained by using the bounds (\ref{b1}) and
(\ref{b2}) for scenario I and scenarios (II,III) respectively.  It should
be emphasized that these bounds are rather conservative as they take only
into account the RGE bound in $\Im\Lambda_t$ (through $\klpn$) and the
bound on $\Re\Lambda_t$ from (\ref{KLSD}).  On the other hand, if
$\Lambda_t$ is almost purely imaginary, as required by the RGE constraints
for a large $\Im \Lambda_t$, the upper bound on $\Re \Lambda_t$ is
generally stronger than the one from (\ref{KLSD}) and one has milder
enhancements of ${\rm BR}(K^+ \rightarrow \pi^+ \nu \bar\nu)$ than in the
``*" case. That is, in order to find the true bound, the correlation
between $\Im \Lambda_t$ and $\Re \Lambda_t$ through RGE should be taken
into account.
In order to investigate this correlation we have repeated the
analysis for $\kppn$ imposing instead of (\ref{RG}) the
more general RGE constraints
\be\label{RG1}
|\Lambda_t|\le 5.0 \cdot 10^{-4}~, \qquad
|\Re\Lambda_t \Im\Lambda_t|\le 0.8\cdot 10^{-9}~,
\ee
derived from (\ref{rgb2}-\ref{rgb3b}).
The results of this analysis are represented by 
${\rm BR}(\kppn)$ without ``*" in tables. As expected the bounds are
stronger than previously obtained. Moreover the sensitivity to $\epp$
diminished and the bounds are mainly governed by $\kmm$ and RGE. 

\subsubsection{Scenario C}
Within this scenario it is possible, in principle, to have a partial
cancellation of the SUSY contributions to $\epp$ generated by $Z$-penguin
and chromomagnetic operators.  Given the strong RGE bound (\ref{RG}), this
possibility has only a minor impact on the upper bounds of both ${\rm
BR}(\klpn)$ and ${\rm BR}(\kppn)$, with respect to scenario B. The only
difference is that a sizable enhancement can also occur for $\Im \Lambda_t
> 0$, if $R_g \Im \Lambda^-_g$ is positive and compensate for the negative
contribution to $(\epp)$ generated by the $Z$ penguin. This would then
allow large values of $K\to\pi\nu\bar{\nu}$ widths also within scenario
I. This case is shown in Table~\ref{tab:rarepo6}.  As can be noticed,
the upper bounds for the two neutrino modes within scenario II and III are
the same as in Table~\ref{tab:rarepo3} (with $\Re(\epp)\leq 30.4 \cdot
10^{-4}$) but, as anticipated, sizable enhancements occur also within
scenario I. Due to the additional independent SUSY contribution to $\epp$,
in all cases (I-III) the upper bounds of $K\to\pi\nu\bar{\nu}$ widths are
insensitive to the experimental constraints on $\epp$ and depend only on
the maximal value of $\lambda_t$.

\begin{table}[t]
\begin{center}
  \begin{tabular}{|l|c|c|c|c|}
  \hline 
  \multicolumn{1}{|c}{Scenario for $\lambda_t$:}
  & \multicolumn{1}{c}{I} & \multicolumn{1}{c}{II} & III  & SM  \\ \hline 
  $10^{10}~{\rm BR}(\klpn)$   & $3.8~(3.8)$ &$1.7~(1.7)$ & $4.0~(4.0)$
          & 0.4 \\ \hline 
  $10^{10}~{\rm BR}(\kppn)^*$ & $2.6~(2.6)$ &$2.4~(2.4)$ & $2.9~(2.9)$
          & 1.1 \\
  $10^{10}~{\rm BR}(\kppn)$   & $1.8~(1.8)$ &$2.1~(2.1)$ & $2.7~(2.7)$
          & 1.1  \\ \hline 
  $10^{11}~{\rm BR}(\kpe)_{\rm dir}\quad [+]$
          & $10.0~(9.3)$ & $5.7~(5.3)$ & $10.3~(9.7)$ & 0.7  \\ 
  $10^{11}~{\rm BR}(\kpe)_{\rm dir}\quad [-]$
          & $5.7~(5.5)$ &$4.9~(4.5)$ & $6.8~(6.1)$ & 0.7  \\ \hline 
  \end{tabular}
\end{center}
  \caption{Upper bounds for the branching ratios 
   of the rare decays $\klpn$, $\kpe$ and $\kppn$ in scenario C, for
   $\Im\Lambda_t>0$ and $R_g \Im \Lambda^-_g >0$, imposing $\Re(\epp)\leq
   30.4(20.0) \cdot 10^{-4}$.  The results in the last two lines are
   obtained for $(\Im \Lambda_g^+/\Im \Lambda_g^-) B_T/(B_G \sqrt{R_S})=
   \pm 1$. For an explanation of the $^*$ see caption 
   of Table \protect{\ref{tab:rarepo1}}.
   \label{tab:rarepo6}}
\end{table}

More interesting is the case of $\kpe$, sensitive to both $\Im\Lambda_t$
and the SUSY contribution to magnetic operators.  Also in this mode the
largest enhancements occur when both $\Im \Lambda_t$ and $R_g \Im
\Lambda^-_g$ are positive, so that $|R_g \Im \Lambda^-_g|$ can reach its
maximum value.  As shown in Table~\ref{tab:rarepo6}, in this case one can
reach values of ${\rm BR}(\kpe)_{\rm dir}$ larger than in scenarios A and
B.  An evidence of ${\rm BR}(\kpe)_{\rm dir} \gsim 10^{-10}$ would provide
a clear signature of this particular (though improbable) configuration.

We finally note that, within scenario C, by relaxing the RGE bound
(\ref{RG}) it is possible to recover the maximal enhancements for the rare
decays pointed out in \cite{CI}. Needless to say, this possibility is
rather remote, as it requires a few fine-tuning adjustments. However it is
interesting to note that in the near future it could be excluded in a truly
model-independent way by more stringent bounds on ${\rm BR}(\kppn)$.
Indeed if ${\rm BR}(\klpn) > 2 \cdot 10^{-9}$ one expects from isospin
analysis \cite{isospin} that ${\rm BR}(\kppn) > 4.6 \cdot 10^{-10}$, not
far from the recent upper bound on this mode obtained by BNL-E787
\cite{E787}.

\section{Summary}
In this paper we have analyzed the rare kaon decays $\klpn$, $\kppn$,
$\kpe$ and the CP violating ratio $\epp$ in a general class of
supersymmetric models. We have argued that only dimension-4 and 5 operators
may escape the phenomenological bounds coming from $\Delta S=2$ transitions
and contribute substantially to $\Delta S=1$ amplitudes. On this basis we
have introduced three effective couplings which characterize these
supersymmetric contributions: $\Lambda_t$ for the $Z$ penguin and
$\Lambda_g^\pm$ for the magnetic ones.  $\Im \Lambda_t$ enters all rare
decays and $\epp$, $\Im\Lambda_g^-$ only $\epp$ while $\Im\Lambda_g^+$ only
$\kpe$. $\Re\Lambda_t$ is important for $\kppn$ and $\kmm$. Since
$\Im\Lambda_g^-$ and $\Im\Lambda_g^+$ are expected to be similar in
magnitude, a connection between $\epp$ and $\kpe$ follows in models with
small $\Im\Lambda_t$.

We have demonstrated explicitly that
\begin{itemize}
\item
the size of $\Im\Lambda_g^\pm$ is dominantly restricted by the
present experimental range of $\epp$;
\item
the size of $\Im\Lambda_t>0$ is bounded by the minimal value of
$\epp$;
\item
the size of $\Im\Lambda_t<0$ is bounded by the renormalization
group analysis (RGE) combined with the experimental values 
on $\ep$ and $\Delta M_K$; 
\item
the size of $\Re\Lambda_t$ is bounded by $\kmm$ and RGE.
\end{itemize}
The imposition of the RGE bounds on the effective couplings has a
considerable impact on the upper bounds of rare kaon decays (e.g. compare
Table~\ref{tab:rarepo5} to Tables~\ref{tab:rarepo1} and \ref{tab:rarepo3})
so that the maximal branching ratios are found to be substantially lower
than those obtained in \cite{CI,BS98}. Given the important role of this
bound it is worth emphasizing that it requires more theoretical 
input than the low-energy phenomenological bounds usually taken 
into account within the mass-insertion approximation. 
Indeed it requires a control on the degrees of freedom of 
the theory up to scales of the order of $10^{16}$ GeV.

In order to accurately describe the relations between $\epp$ and the rare
decays we have performed a numerical analysis in three basic scenarios:
\begin{description}
\item{{\bf Scenario A}: $[\Im\Lambda_t=0,\ \Im\Lambda_g^\pm \not=0]$.}
\item{{\bf Scenario B}: $[\Im\Lambda_t\not=0,\ \Im\Lambda_g^\pm=0]$.}
\item{{\bf Scenario C}: $[\Im\Lambda_t\not=0,\ \Im\Lambda_g^\pm\not=0]$.}
\end{description}
In each of these scenarios we have considered three scenarios
for the CKM factor $\lambda_t$:
\begin{description}
\item{{\bf Scenario I}: $\lambda_t$ is taken from the standard analysis of
the unitarity triangle.}
\item{{\bf Scenario II}: $\Im\lambda_t=0$ and $\Re\lambda_t$ is varied
            in the full range consistent with the unitarity
            of the CKM matrix.}
\item{{\bf Scenario III}: $\lambda_t$ is varied
            in the full range consistent with the unitarity
            of the CKM matrix.}
\end{description}
As we have discussed, scenario A with scenarios I or II for
the CKM matrix is most natural within supersymmetric models with
approximate flavour symmetries. However the other scenarios cannot
be excluded at present and we have analyzed them in detail.
Our main findings, collected in
Tables~\ref{tab:KLee}-\ref{tab:rarepo3} 
and \ref{tab:rarepo6}
are as follows:
\begin{itemize}
\item
In scenario A there is room for enhancement of ${\rm BR}(\kpe)_{\rm dir}$ 
by up to one order of magnitude and of ${\rm BR}(\kppn)$ by factors
2-3 over the Standard Model expectations. ${\rm BR}(\klpn)$ remains generally
in the ball park of the Standard Model expectations 
except for scenario II,
where it becomes vanishingly small.
\item 
In scenario B, with the Standard Model values of $\Im\lambda_t$ (I),
enhancements of ${\rm BR}(\klpn)$ by factors 2-3 and of ${\rm
BR}(\kpe)_{\rm dir}$ by factors 3-5 are still possible, while ${\rm
BR}(\kppn)$ can be enhanced by at most a factor of 2.  On the other hand, in
scenarios II and III enhancements of ${\rm BR}(\kpe)_{\rm dir}$ and ${\rm
BR}(\klpn)$ by one order of magnitude and of ${\rm BR}(\kppn)$ up to a
factor of 3 over Standard Model expectations are possible.
These upper limits are dictated by the RGE bounds.
\item 
In scenario C enhancements of rare-decay branching ratios larger than
in scenarios A and B are only possible if $\Im \Lambda_g^-$ and
$\Im\Lambda_t$ have the same sign so that the contributions of the
chromomagnetic penguin and $Z^0$-penguin to $\epp$ cancel each other to
some extent. As a consequence the restrictions from $\epp$ are
substantially weakened and what matters are the RGE constraints. 
In this rather improbable scenario one order of magnitude 
enhancements of ${\rm BR}(\klpn)$ are possible even 
if the standard determination of $\lambda_t$ is valid
and ${\rm BR}(\kpe)_{\rm dir}$ could reach the 
$10^{-10}$ level. On the
other hand ${\rm BR}(\kppn)$, being mainly sensitive to $\Re\lambda_t$ and
$\Re\Lambda_t$, stays always below $3\cdot 10^{-10}$ 
as in scenarios A and B.
\end{itemize}

We observe certain patterns in each scenario which will allow to
distinguish between them, and possibly rule them out once data on rare
decays and improved data and theory for $\epp$ will be available. In
particular in the near future with more stringent bounds on ${\rm
BR}(\kppn)$ the most optimistic enhancements 
(like those occurring in scenarios C or B.III)
could be considerably constrained. 
In the more distant future, a clean picture will emerge 
from the measurements of ${\rm BR}(\klpn)$ and 
${\rm BR}(\kpe)_{\rm dir}$.

\vfill

\section*{Acknowledgments}
A.J.B and L.S. have been partially supported
by the German Bundesministerium f\"ur
Bildung und Forschung under contracts 06 TM 874 and 05 HT9WOA. 
G.C. and G.I have been partially supported
the TMR Network under the EEC Contract  
No. ERBFMRX--CT980169 (EURODA$\Phi$NE).
The work of A.R. was supported by the TMR Network under the EEC
Contract  No. ERBFMRX--CT960090.

\end{document}